\documentclass[longauth]{aaEC}

\usepackage{graphicx}
\usepackage{natbib}
\usepackage{scalerel}
\usepackage{siunitx}
\usepackage{lastpage}
\usepackage{soul}
\usepackage{amsmath}
\usepackage{txfonts}

\usepackage[table]{xcolor}

\newcolumntype{P}[1]{>{\centering\arraybackslash}p{#1}}
\newcommand*{\rom}[1]{\expandafter\@slowromancap\romannumeral #1@}
\newcommand{\orcid}[1]{} 

\usepackage{txfonts}
\usepackage[pdfencoding=auto,psdextra]{hyperref}
\hypersetup{
    colorlinks=true,
    linkcolor=blue,
    filecolor=magenta,      
    urlcolor=blue,
    citecolor=blue
}
\def\Tab{\mbox{Table~}}

\makeatletter
\renewcommand*\aa@pageof{, page \thepage{} of \pageref*{LastPage}}
\makeatother

\usepackage[utf8]{inputenc}
\usepackage{datetime2}

\usepackage[switch, modulo]{lineno}

\usepackage{euclid}

\begin{document}

%
%

\title{The quasi-star model for little red dots: Potential and challenges}

\renewcommand{\orcid}[1]{} 
\author{Fabrizio Gentile\thanks{\email{fabrizio.gentile@cea.fr}}\inst{1,2}
\and
Mauro Giavalisco\inst{3}
\and 
Emanuele Daddi\inst{1}
\and 
David Elbaz\inst{1}
\and 
Jean-Baptiste Billand\inst{1}
\and
Maximilen Franco\inst{1}
\and
Benjamin Magnelli\inst{1}
\and 
Guillermo Barro\inst{4}
\and
Yingjie Cheng\inst{5}
\and
Nikko J. Cleri\inst{6,7,8,}
\and
Kelcey Davis\thanks{NSF Graduate Research Fellow}\inst{9,10}
\and
Ivan Delvecchio\inst{2}
\and
Mark Dickinson\inst{11}
\and
Steven L. Finkelstein\inst{12}
\and
Giovanni Gandolfi\inst{13,14}
\and
Michaela Hirschmann\inst{15}
\and
Weida Hu\inst{16,17}
\and
Jeyhan S. Kartaltepe\inst{18}
\and
Dale Kocevski\inst{19}
\and
Anton M. Koekemoer\inst{20}
\and
Ray A. Lucas\inst{20}
\and
Sara Mascia\inst{21}
\and
Lorenzo Napolitano\inst{22}
\and
Casey Papovich\inst{16,17}
\and
Borja Pérez-Díaz\inst{23}
\and
Pablo Perez-Gonzalez\inst{24}
\and
Jonathan R. Trump\inst{9}
\and
Xin Wang\inst{25,26,27}
\and
L. Y. Aaron Yung\inst{20}
}

\institute{Université Paris-Saclay, Université Paris Cité, CEA, CNRS, AIM, 91191 Gif-sur-Yvette, France
\and
INAF -- Osservatorio di Astrofisica e Scienza dello Spazio di Bologna, Via Piero Gobetti 93/3, 40129 Bologna, Italy
\and
University of Massachusetts Amherst, 710 North Pleasant Street, Amherst, MA 01003-9305, USA
\and
University of the Pacific, Stockton, CA 90340, USA
\and
Department of Astronomy, University of Washington, Seattle, WA, 98195, USA
\and
Department of Astronomy and Astrophysics, The Pennsylvania State University, University Park, PA 16802, USA
\and
Institute for Computational and Data Sciences, The Pennsylvania State University, University Park, PA 16802, USA
\and
Institute for Gravitation and the Cosmos, The Pennsylvania State University, University Park, PA 16802, USA
\and
Department of Physics, 196A Auditorium Road, Unit 3046, University of Connecticut, Storrs, CT 06269, USA
\and
Los Alamos National Laboratory, Los Alamos, NM 87545, USA
\and
NSF National Optical-Infrared Astronomy Research Laboratory, 950 North Cherry Ave., Tucson, AZ 85719, USA
\and
Department of Astronomy, The University of Texas at Austin, Austin, TX 78712, USA
\and
Dipartimento di Fisica e Astronomia "G. Galilei", Università di Padova, Vicolo dell'Osservatorio 3, 35131, Padova, Italy
\and
INAF -- Osservatorio Astronomico di Padova, Vicolo dell'Osservatorio 5, 35122, Padova, Italy
\and
Institute of Physics, Laboratory of Galaxy Evolution, Ecole Polytechnique F´ed´erale de Lausanne (EPFL), Observatoire de Sauverny, 1290 Versoix, Switzerland
\and
Department of Physics and Astronomy, Texas A\&M University, College Station, TX 77843-4242, USA
\and
George P. and Cynthia Woods Mitchell Institute for Fundamental Physics and Astronomy, Texas A\&M University, College Station, TX
\and
Laboratory for Multiwavelength Astrophysics, School of Physics and Astronomy, Rochester Institute of Technology, 84 Lomb Memorial Dr., Rochester, NY 14623, USA
\and
Department of Physics and Astronomy, Colby College, Waterville, ME 04901, USA
\and
Space Telescope Science Institute, 3700 San Martin Drive, Baltimore, MD 21218, USA
\and
Institute of Science and Technology Austria (ISTA), Am Campus 1, 3400 Klosterneuburg, Austria
\and
INAF -- Osservatorio Astronomico di Roma, via Frascati 33, 00078, Monteporzio Catone, Italy
\and
Dipartimento di Fisica, Università di Roma Sapienza, Città Universitaria di Roma -- Sapienza, Piazzale Aldo Moro, 2, 00185, Roma, Italy
\and
Centro de Astrobiologia (CAB), CSIC-INTA, Ctra. de Ajalvir km 4, Torrej´on de Ardoz, E-28850, Madrid, Spain
\and
School of Astronomy and Space Science, University of Chinese Academy of Sciences (UCAS), Beijing 100049, China
\and
National Astronomical Observatories, Chinese Academy of Sciences, Beijing 100101, China
\and
Institute for Frontiers in Astronomy and Astrophysics, Beijing Normal University, Beijing 102206, China
}

 \date{\today}

\abstract{Little red dots (LRDs) are a class of sources discovered by the \textit{James Webb} Space Telescope observationally defined by a V-shaped rest-frame UV-optical spectral energy distribution, a compact or unresolved morphology, and, frequently, broad hydrogen emission lines. How these characteristics translate into physical and structural properties remains unclear. Among various models, those involving a quasi-star interpret LRDs as an intermediate stage in the evolution of a super-massive black hole (SMBH) seed into a classic active galactic nucleus. In this paper, we employ the radiative-transfer code \texttt{Cloudy} to study whether this model is able to reproduce the spectral features commonly observed in LRDs. The model consists of an accreting SMBH ($M_{\rm BH}\sim10^{5-6} \ M_\odot$) surrounded by a convective layer where a black-body (BB) spectrum with $T\sim5000 \ {\rm K}$ and $L\sim10^{44.4} \ {\rm erg \ s}^{-1}$ is produced. This BB spectrum is then reprocessed by a concentric thick ($\Delta R\sim1000 \ {\rm AU}$) shell of dense ($n_{\rm H}\sim10^{11} \ {\rm cm}^{-3}$) gas partially ionised by thermal collisions. The emerging radiation is further reprocessed by a diffuse clumpy medium surrounding the quasi-star. We fit this model to JWST/NIRSpec spectra of LRDs from the literature, deriving the main physical parameters and the SMBH masses. Once coupled with the UV emission from a host galaxy, this model is able to reproduce the shape of the UV-to-NIR continuum, including the presence of a Balmer break, as well as the luminosity of the hydrogen emission lines. However, this quasi-star model does not natively account for the presence of broad helium lines and for the possible presence of hot dust, needing additional components to match these observables. Our main result is to show how some LRDs can be modelled as quasi-stars, highlighting that a significant degeneracy exists among different LRD models. This has important consequences for our understanding of the mechanisms driving black hole growth in the early Universe.}


   \titlerunning{Quasi-stars and LRDs: Potential and challenges}
   \authorrunning{F. Gentile et al.}
   
   \maketitle

\section{Introduction}
\label{sec:intro}

One of the main results of the first years of operation of the \textit{James Webb} Space Telescope (JWST) is the discovery of a previously unidentified population of objects called ‘little red dots’ (LRDs; \citealt{Matthee_24}). These objects were observed for the first time by \citet{Labbe_23} in the Cosmic Evolution Early Release Science Survey (CEERS; \citealt{Finkelstein_25}), appearing as extremely red and compact sources. Since then, LRDs have been selected in most of the extra-galactic surveys conducted with JWST (e.g. \citealt{Akins_25,Labbe_25,Kocevski_25,Barro_26}). Even though the selection criteria adopted to collect the sources vary, a consensus has emerged on the definition of LRDs as objects that satisfy three main observational criteria. Firstly, LRDs are extremely compact in the rest-frame optical and near-infrared (NIR), resulting in unresolved or partially resolved images in the red filters of the JWST/NIRCam instrument (F277W and F444W; see e.g. \citealt{Akins_25,Hviding_25}). Secondly, their rest-frame UV/optical spectral energy distribution (SED) presents a characteristic ‘V’ shape, with a blue UV arm and a red optical one (e.g. \citealt{Kocevski_23,Labbe_25}), and with the apex of the V located -- in the vast majority of cases --  around the wavelength of the Balmer limit ($\lambda\sim 3645 \AA$; \citealt{Setton_25}). Finally, LRDs are characterised by the presence of broad hydrogen emission lines, even though this last property is not part of the selection criteria but rather the result of follow-up spectroscopic studies \citep[e.g.][]{Kocevski_23,Greene_24,Matthee_24,Hviding_25,Taylor_25}. Building on these criteria, several samples of LRDs have been assembled in the past few years, enabling  population-level studies aimed at characterising their overall properties. Key results from these investigations include extreme faintness at (sub-)millimetre, radio, and X-ray wavelengths (e.g. \citealt{Casey_24,Ananna_24,Akins_25,Yue_24,Kocevski_25}), and slow or absent variability in the UV and X-rays (e.g. \citealt{Zhang_25,Kokubo_25}). Finally, LRDs are mainly found at $z>4$, with a steep decrease in their number density at lower redshifts \citep[e.g.][]{Kocevski_25,Ma_26}. This last property raises important questions about the identification of the descendants of these objects \citep[e.g.][]{Billand_25,Chen_26}.

The criteria adopted to select LRDs are purely observational and cannot immediately be connected to a specific physical property. Thus, there is growing interest in understanding whether the observed properties of LRDs can be accounted for by a single physical model, and what the nature of that model would be. For example, an early interpretation of the red continuum as due to dust-obscured stellar populations (e.g. \citealt{Labbe_23,Baggen_24,Perez-Gonzalez_24}) was found to be in strong tension with the stringent upper limits on the dust content of these sources from far-infrared (FIR) observations (e.g. \citealt{Casey_24,Akins_25,Xiao_25,Setton_25b}). On the other hand, simple models including an active galactic nucleus (AGN) as the main driver of the optical-NIR emission are often unable to reproduce the observed shape of the SED in the rest NIR (e.g. \citealt{Perez-Gonzalez_24,Williams_24}, but see different conclusions based on stacking analysis by \citealt{Delvecchio_25} and \citealt{PerezGonzalez_26}). Additionally, dynamical estimates of the mass of the super-massive black holes (SMBHs) at the centre of these objects are often in tension with current models of black hole growth, with ‘over-massive black holes’ presenting a ratio with the stellar mass of the host galaxy up to 1000 times higher than the local relations (see e.g. \citealt{Pacucci_23,Chen_25a,Jones_25,Maiolino_25,Tripodi_25,Greene_26}).

To overcome these challenges, other more exotic models have been explored to explain the nature of LRDs. Some of these models introduced comparatively new scenarios, such as primordial black holes (PBHs; \citealt{Dayal_26}), nuclear star-clusters \citep{Kritos_25}, direct-collapse black holes \citep{Pacucci_26}, progenitors of globular clusters \citep{Chisholm_26}, super-massive stars (e.g. \citealt{Martins_26}), or AGN surrounded by a dense layer of gas \citep{Ji_25,Madau_26}. Other models revisited older ideas such as the so-called quasi-stars, in which an accreting SMBH is surrounded by a dense atmosphere of gas that fully thermalises the radiation, analogously to a star. These objects were initially proposed by \citet{Begelman_08} as a possible mechanism able to explain the early growth of SMBHs in the high-\textit{z} Universe, and then revisited to explain the key properties of LRDs after their discovery by JWST \citep{Begelman_26}. 

Models conceptually similar to the quasi-star by \citet{Begelman_08}, dubbed black-hole stars (BH*s), have been recently proposed to explain the extremely large Balmer breaks observed in some LRDs (e.g. \citealt{DeGraaff_25a,Naidu_25,Taylor_25,Ronayne_26}), which are very difficult to reproduce with standard stellar population models and known dust extinction laws. The physical structures of the quasi-star and the BH* are similar, since both models have an accreting SMBH surrounded by a dense distribution of gas. However, the key difference consists in the shape of the radiation field emerging from the dense gas. In the quasi-star by \citet{Begelman_08}, the SMBH is surrounded by a zone of ‘saturated convection’, where the radiation emitted from the accretion disk is completely thermalised as it propagates in the surrounding layers, and thus assumes the shape of a black-body (BB) spectrum. In the BH* (following the implementation by \citealt{Naidu_25}), no convective zone is present and the dense atmosphere is directly irradiated by the emission of the accretion disk, without full thermalisation.

Observational evidence of thermalised radiation (potentially indicating the presence of a convective envelope) has been recently suggested by \citet{Kido_25,DeGraaff_25b,Sun_26} based on the shape of the continuum of a large sample of LRDs at wavelengths redder than the Balmer break, which can be empirically modelled with a modified black body with $T_{\rm BB}\sim5000 \ {\rm K}$. At the moment, these models do not include any treatment of the emission lines (i.e. their production and transfer), or of the Balmer break. It is therefore unclear how the lines can form in such a scenario, namely if they are created by photoionisation in the proximity of the SMBH and survive radiation transfer or in the external layers. In the former case, it is unclear how these emission lines can get out of the convective envelope without being thermalised. In the latter case, one needs to address the process responsible for the formation and broadening of the lines, given the lack of ionising radiation in a BB with $T\sim5000 \ {\rm K}$ and the limited effect of the SMBH gravity at large distances. In this paper, we revisit the model of the quasi-star by \citet{Begelman_08} to address the question on the shape of the continuum radiation of LRD, including the Balmer break, and the production and luminosity of the hydrogen emission lines thanks to a new suite of radiative-transfer simulations performed with \texttt{Cloudy} (version C25; \citealt{Ferland_98,Cloudy_25}) and to the comparison with a large sample of LRDs with high-quality NIRSpec/PRISM data. 

The paper follows this structure. In Sect. \ref{sec:methods} we present the \texttt{Cloudy} modelling of the quasi-star and the relation between its free parameters and the observational properties of LRDs. In Sect. \ref{sec:results} we apply the model to real spectroscopic data of LRDs observed with JWST and fit the values of the main free parameters. In Sect. \ref{sec:discussion} we discuss our results and compare them with alternative models of LRDs. Finally, we draw our conclusions in Sect. \ref{sec:summary}. Throughout this paper, we assume a standard $\Lambda$ cold dark matter cosmology with the parameters reported in \citet{Planck_18} and a \citet{Chabrier_03} initial mass function.

\section{A model for little red dots}
\label{sec:methods}

\begin{figure*}
    \centering
    \includegraphics[width=\linewidth]{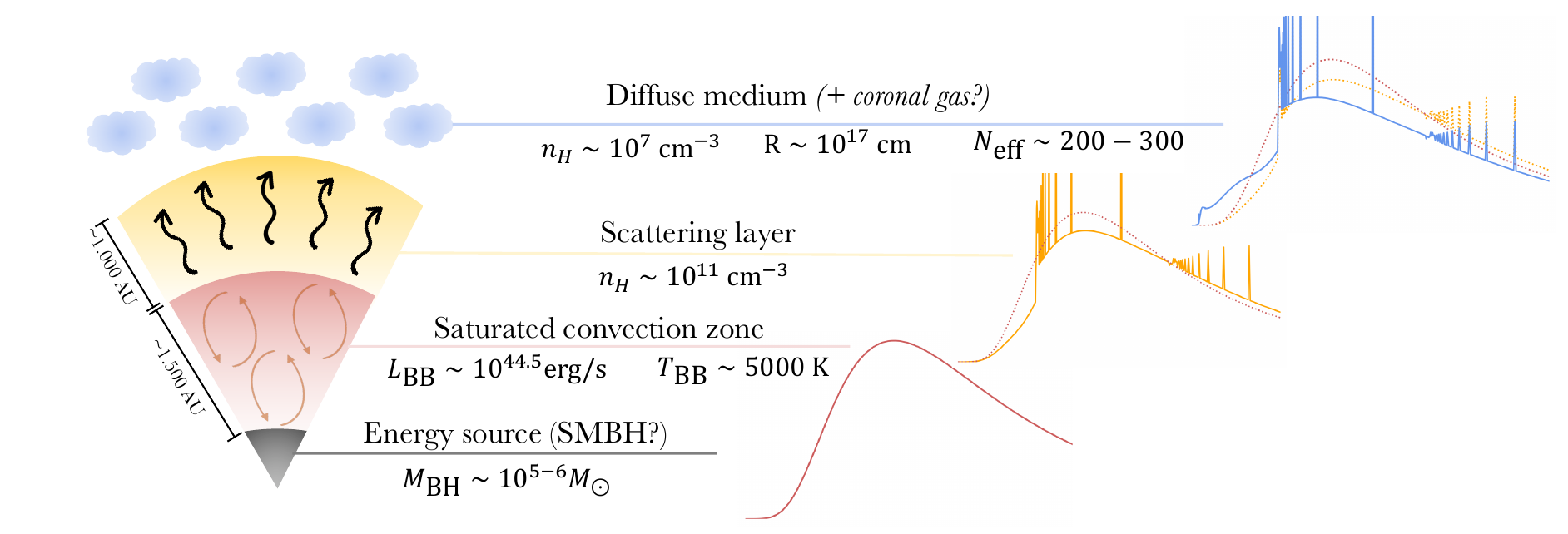}
    \caption{Sketch representing the main model given in input to \texttt{Cloudy}. Our quasi-star model is inspired by \citet{Begelman_08} and includes a source of energy surrounded by a convective shell and by a dense layer of gas. The quasi-star is then surrounded by a diffuse and clumpy medium, with the possible presence of a warmer corona (see Sect. \ref{sec:He_lines}). The spectrum produced by each layer is shown on the right part of the figure. The quantities reported are those obtained by fitting the model to the sample of LRDs from \citet{DeGraaff_25b}.}
    \label{fig:skectch}
\end{figure*}

\subsection{Cloudy model of a quasi-star}
\label{sec:model}

Our \texttt{Cloudy} setup implements the original model of the quasi-star presented in \citet{Begelman_08} and \citet{Begelman_26}. In this model (see Fig. \ref{fig:skectch} for a visual representation), the central source of energy (e.g. an accreting black hole) is completely surrounded by a convective and optically thick envelope. In the quasi-star model, the density and thickness of this envelope are assumed high enough that the radiation becomes completely thermalised at all wavelengths. Hence, the spectrum of the radiation that emerges from this region is a BB characterised by two free parameters: its temperature ($T_{\rm BB}$) and total luminosity ($L_{\rm BB}$). This convective zone is then surrounded by a shell of dense gas (the ‘scattering layer’, hereafter), where radiative transfer of the BB radiation takes place. The physical processes taking place in this shell modify the shape of the input spectrum and allow for the formation of emission lines. To obtain results more easily comparable with other studies, we assumed a uniform hydrogen density ($n_H$) within the scattering layer. We highlight that this is likely an over simplification; however, we have directly verified that adopting a more realistic profile, such as a power law (with the slope being a free parameter), does not significantly affect our results, except for the physical scale of the layer (see Sect. \ref{sec:properties_results} for a more detailed discussion). We also assumed that this shell of gas is in direct contact with the convective zone. Therefore, the outer radius of the convective shell, $R_{BB}$, can be fixed to the inner radius of the scattering layer, which is linked to the properties of the black body through the Stefan-Boltzmann law
\begin{equation}
\label{eq:boltzmann}
    R_0\equiv R_{BB}=\sqrt{\frac{L_{\rm BB{}}}{4 \pi \sigma T_{\rm BB}^4}} \ ,
\end{equation}
where $\sigma$ is the Stefan-Boltzmann constant. With this choice, the shell is completely described by two free parameters: $n_{\rm H}$ and $\Delta R$, the thickness of the shell. The grid of parameters employed in our modelling is reported in \Tab\ref{tab:params}.

Differently from the original model by \citet{Begelman_08}, we included a final additional component to our setup of the \texttt{Cloudy} simulation, consisting in a series of scattered clouds surrounding the quasi-star. In our implementation, we assumed a set of identical clouds, with an open geometry characterised by a uniform gas density of $n_{\rm H,c}=10^7 \ {\rm cm}^{-3}$ and a fixed thickness of $\Delta R_{c}=10^{17} \ {\rm cm}$, and located at a constant distance of $10^{15} \ {\rm cm}$ from the external border of the quasi-star. Since these parameters are found to be mostly degenerate with the geometrical setup of the clouds, we chose them to resemble the usual properties of the clouds in the broad-line region (BLR) of classical AGN (see e.g. \citealt{Osterbrock_06} and references therein). Since the goal of this study is to verify whether this model is able to reproduce the shape of the continuum and the luminosity of the hydrogen emission lines, for simplicity we adopted primordial abundances for both the scattering layer and the surrounding clouds, reducing the size of the parameter space from that required to explore metal abundances and emission lines. This assumption should be in principle in agreement with the original model by \citet{Begelman_08}, presenting the quasi-star as a late evolutionary stage of a massive star, even though a small amount of metals could be expected in the inner regions of the system thanks to the nucleosynthesis taking place before the formation of the quasi-star (see Sect. \ref{sec:dust}).

To model the overall geometrical setup of the clouds, we made the following simple approximation. In the standard \texttt{Cloudy} output, the radiation reprocessed by a cloud is divided into three different components. The first is the transmitted attenuated continuum observed from the external face of the cloud relative to the incoming radiation; the second is the reflected radiation, which comes from the illuminated face; the last is the isotropically diffused radiation by the cloud. If the covering factor of the clouds is less than unity, all these components contribute to the observed spectrum in a manner that depends on the spatial displacement of the clouds. For simplicity, we modelled the geometry of the clouds by multiplying the reflected and diffused spectra by two free parameters, $\beta$ and $\gamma$, respectively, which can assume both positive and negative values to account for (positive) contributions due to reflections into the line of sight or (negative) absorption from the line of sight. With this approximation, the parameters $\beta$ and $\gamma$ are related to the effective number of clouds ($N_{\rm eff}$; i.e.  also account for inclination and the contribution from other lines of sight). We acknowledge that this linear approximation represents an over-simplification, since it does not fully capture the case of a chain of clouds, each of which sees a different spectrum from that emerging from the scattering layer due to the transfer through previous clouds in the chain. We have verified, however, that the reflected and diffused spectra represent less than 0.1\% of the incoming spectrum. Therefore, we consider this approximation valid until the two $\beta$ and $\gamma$ parameters are smaller than $10^3$.

Finally, we note that -- due to the characteristics of \texttt{Cloudy} -- our model does not capture the kinematics of the gas, with the only exception of a turbulence factor of $\Delta v=500 {\rm km \ s}^{-1}$ included in the scattering layer and in the clouds for consistency with \citet{Naidu_25}, where this value is found to match the width of the absorption features in the Balmer lines of LRDs. As a consequence, our model is not able to predict the profile of the emission lines produced in the quasi-star nor to account for small redshift effects due to the presence of inflowing or outflowing gas (see e.g. \citealt{Sneppen_26}).

\subsection{Link between physical and observational properties of LRDs}
\label{sec:properties}

The full \texttt{Cloudy} model is determined by a set of six free parameters. Four of them related to the quasi-star ($L_{\rm BB},T_{\rm BB},n_{\rm H}$, and $\Delta R$) and two to the number and geometry of the surrounding clouds ($\beta$ and $\gamma$). Before applying this model to real data, we focus on how these parameters affect the output spectrum.

\paragraph{Scattering layer:} As in the theoretical model by \citet{Begelman_08,Begelman_26}, the shell of dense gas is responsible for the formation of the Balmer break and of the hydrogen emission lines (see Fig. \ref{fig:skectch}). The output models show a significant correlation between the integrated hydrogen column density
\begin{equation}
    N_H=\int_{R_0}^{R} n_H(r) \ {\rm d}r \ ,
\end{equation}
and the presence of stronger breaks and brighter lines. 

It is interesting to notice that at normal gas densities, such as those encountered in HII regions, the ionising fraction in presence of a black body with $T_{\rm BB}\sim5000 - 7000 \ K$ would be effectively null, since such radiation cannot ionise hydrogen due to the lack of ionising photons. However, at high densities -- such as those reached in the scattering layer in our modelling -- the contribution of thermal collisions becomes predominant (above 99\% in the \texttt{Cloudy} computation), allowing for the formation of emission lines. For the same reason, the $n=2$ level of the hydrogen atoms becomes over-populated with respect to the local thermal equilibrium, causing the formation of the strong Balmer breaks visible in LRDs (e.g. \citealt{DeGraaff_25a,Naidu_25,Taylor_25,Inayoshi_25}). In addition, the collisional ionisation and the over-population of excited levels are also at the origin of the large Balmer decrements that are commonly observed in LRDs (e.g. \citealt{DeGraaff_25b}), since the main assumptions behind case B recombination are not satisfied in these conditions (see e.g. \citealt{Osterbrock_89}). Finally, as discussed in \citet{Begelman_26}, the large column densities of free electrons produced by the collisional ionisation can contribute to the broadening of the lines through repeated electron scattering. In addition, the over-population of the $n=2$ level allows for Balmer line broadening via resonant scattering, as suggested -- for instance -- by \citet{Naidu_25,Chang_25,Matthee_26}. Both effects could in principle justify the correlation observed by \citet{Sneppen_26} and \citet{Matthee_26} between the broadening of the H$\alpha$ line and the strength of the Balmer break. This hypothesis, however, cannot be directly tested with \texttt{Cloudy} which cannot predict the resolved shape of the emission lines.

Another interesting effect of the scattering layer resides in the modification of the spectral shape of the BB continuum as the radiation is transferred. As can be seen in Fig. \ref{fig:skectch}, part of the blue spectrum processed by the shell of gas is re-emitted at longer wavelengths, altering the slope of the continuum relative to the Rayleigh-Jeans tail of the Planck function. This result is particularly interesting because it explains one of the key observational findings by \citet{DeGraaff_25b}, namely that the continuum of LRDs cannot be simply modelled with a ‘classical’ BB. The best-fitting functional form for the vast majority of LRDs is indeed a modified BB obtained by multiplying the Planck function by a factor, $\nu^{\beta}$, modifying the shape of the Rayleigh-Jeans tail (see a more in-depth discussion of this point in Sect. \ref{sec:other_models}).

\paragraph{Diffuse clouds:} The diffuse clouds surrounding the quasi-star have the main effect of changing the continuum shape of the spectrum emitted by the scattering layer. Given the lower hydrogen column density in these clouds (Sect. \ref{sec:model}) and the low temperature of the radiation, the diffuse medium is almost completely neutral. Therefore, it cannot affect the strength of the emission lines nor the Balmer break produced in the scattering layer. However, this neutral gas can act as a source of continuum opacity, through electron and Rayleigh scattering. The first is wavelength-independent and simply causes a change in the overall normalisation of the spectrum. The second is more effective in scattering blue wavelengths, producing a change in the spectral shape in the region surrounding the Balmer break. As an effect, since the open geometry of the clouds allows some of the reflected radiation to come back into the line of sight, the reprocessed spectrum can assume a steeper or broader shape with respect to the original radiation emitted by the scattering layer, depending on whether the dominant contribution comes from, respectively, from the absorbed or reflected emission.

\subsection{Accounting for the host galaxy}
\label{sec:host}

In both the quasi-star and BH* models, the accreting SMBH is mostly responsible for the rest-optical spectrum, at wavelengths redder than the Balmer break (e.g. \citealt{Naidu_25,DeGraaff_25b,Sun_26}). The rest-UV radiation is normally explained as due to young stellar populations belonging to the host galaxy. Several studies pointed out how the UV morphology of a class of LRDs is indeed extended (e.g. \citealt{Rinaldi_25,Ma_26,Cloonan_26}), supporting the non-nuclear origin of the rest-UV spectrum. In this paper, we follow the same assumption and attribute the spectrum at wavelengths bluer than the Balmer break to stellar emission. 

To proceed with modelling of the stellar emission, we simulated a grid of simple stellar populations (SSPs) with the \texttt{Python} implementation of the \texttt{Flexible Stellar Population Synthesis} code (FSPS; \citealt{Conroy_09,Conroy_10}). Since a detailed study of the properties of the host galaxies of LRDs is outside the main goals of this study, we simply simulated a grid of 20 SSPs with stellar ages logarithmically sampled from the range 10 Myr - 1 Gyr and a fixed metallicity of $Z=0.1 \ Z_\odot$\footnote{We emphasise that this value is unconstrained in our model: we fix a reference value and discuss in Sect. \ref{sec:BH_Masses} the impact of this choice on the estimated stellar masses of the host galaxies.}. Each resulting spectrum was normalised to a stellar mass of $M_\ast=1 \ M_\odot$; therefore, we kept the total stellar mass ($M_\ast$) as a free parameter to be fixed during the fitting procedure.

We want to emphasise that, in the majority of cases, including the contribution from the host galaxy is not necessary to model the continuum at wavelengths redder than the Balmer break and the optical hydrogen emission lines. In some cases the quasi-star model even predicts a non-negligible contribution to the spectrum bluer than the break thanks to the Rayleigh scattering of the reprocessed BB by the diffused clouds (see Sect. \ref{sec:properties_results}). The choice of including stellar emission in the fits to explain the full UV spectrum, however, is not without consequences. While the spectrum of the host galaxy becomes negligible compared to that of the quasi-star at wavelengths much longer than the Balmer break, in proximity of the break the two contributions can be comparable, as illustrated in the bottom panel of Figure \ref{fig:ratio}. In such cases fitting the two models together affects the inferred parameters of the quasi-stars and introduces covariances between the parameters of the quasi-star and those of the host galaxy.

\begin{table}[]
    \centering
    \caption{Main parameters given in input to \texttt{Cloudy}.}
    \begin{tabular}{lccc}
        \hline
        \hline
         Parameter & Range & Step & Units  \\
         \hline
         $\log(L_{\rm BB})^{(a)}$ & $42-45$ & 1  & erg s$^{-1}$ \\
         $ T_{\rm BB}$ & $4500-7500$&  500 & K \\
         \hline
         $\log(n_{H,l})$& $9-12$ & 0.5 & cm$^{-3}$ \\
        $\log({\rm \Delta R})$& $14-17.5$ & 0.5 & cm \\
        \hline
    \end{tabular}\\
    {\raggedright $^{(a)}$ {The normalisation of the output model is also affected by the multiplicative $\alpha$ parameter introduced in Sect. \ref{sec:fitting}.} \par}
    \label{tab:params}
\end{table}

\section{Fitting the model to PRISM spectra}
\label{sec:results}

\begin{figure*}
    \centering
    \includegraphics[width=0.49\linewidth]{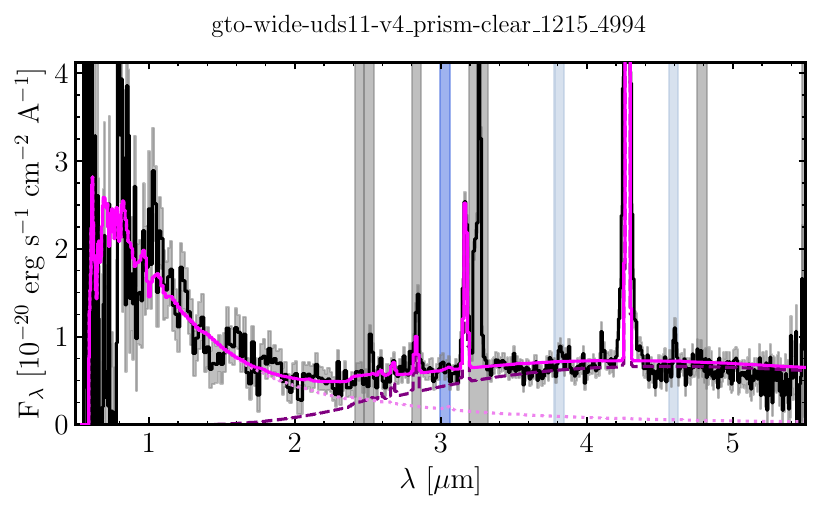} \
    \includegraphics[width=0.49\linewidth]{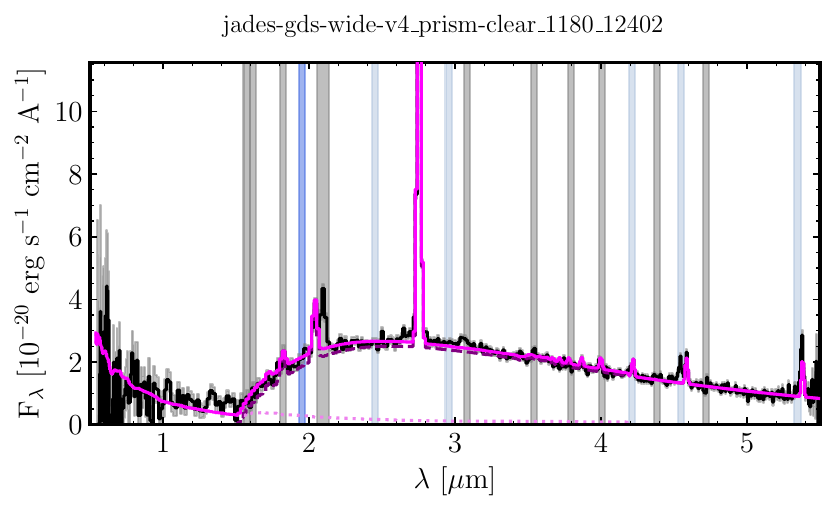} \ 
    \includegraphics[width=0.49\linewidth]{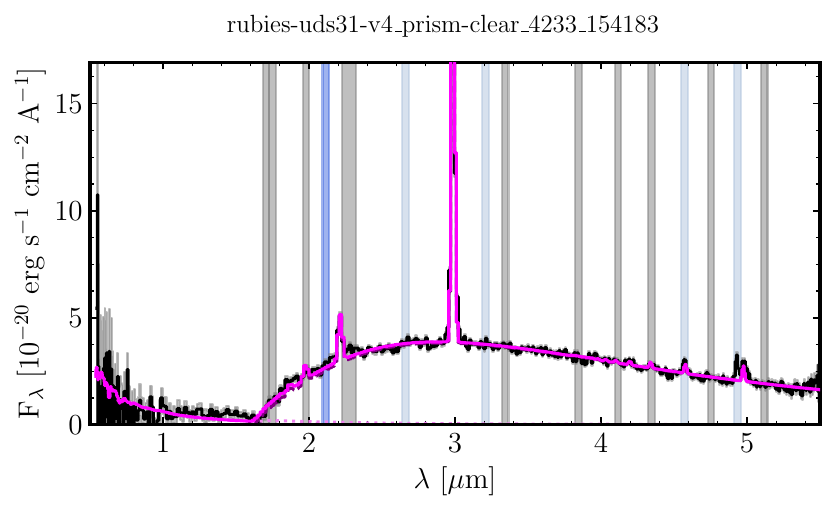} \
    \includegraphics[width=0.49\linewidth]{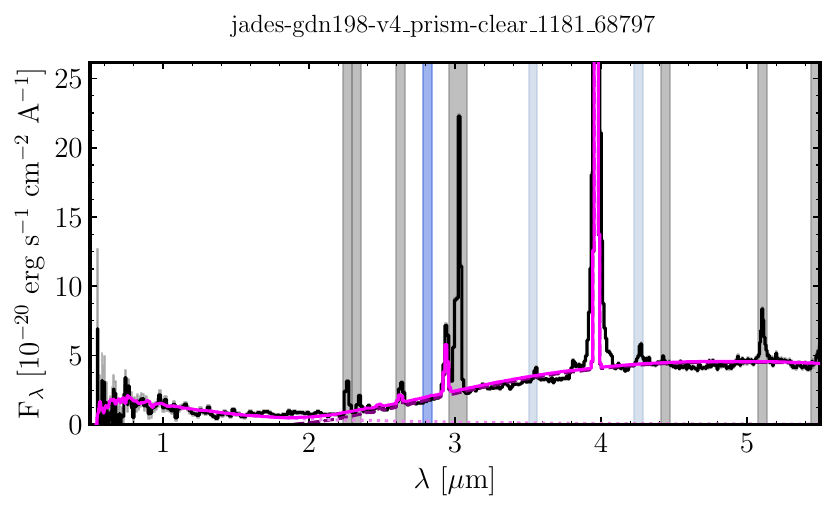} \    
    \caption{Some examples of the quasi-star model applied to LRDs in the sample by \citet{DeGraaff_25b}. The solid black line shows the NIRSpec/PRISM spectrum, while the solid magenta line shows the best-fitting composite model. Its two components are shown as a dashed violet line (quasi-star) and a dotted pink line (host galaxy). The shaded areas indicate the non-hydrogen emission lines that are masked in the fit: metals are indicated in grey, the HeI lines in light blue, and the HeII in darker blue. {These four sources showcase the large variety of spectral shape that are reproducible with the quasi-star model, including different strengths of the Balmer break and contribution of the host galaxy.} The full sample of modelled LRDs can be found in the appendix.}
    \label{fig:examples}
\end{figure*}

In this section we fit the model introduced before to a set of real LRDs from the literature. This test has a double importance. On the one hand, it allows us to fit the free parameters describing our model. On the other hand, it allows us to verify to what extent the quasi-star model is able to reproduce the main observables linked to LRDs. However, it is important to underline how a growing number of studies (see e.g. \citealt{Barro_25,PerezGonzalez_26,Billand_26}) have highlighted how LRDs comprise quite a diverse population of sources, with spectral properties that significantly vary from one object to another. Therefore, it is important to clarify that in this study we do not aim at explaining the whole population of LRDs with a single model. A discussion of the applicability of the quasi-star model to the whole LRD population (as currently selected) and of the limits of such an approach is the subject of a forthcoming paper (Gentile et al., \textit{in prep.}).

\subsection{Spectroscopic and photometric data}
\label{sec:data}

We fitted the quasi-star model to the sample of LRDs compiled by \citet{DeGraaff_25b}. This collection includes 134 objects selected from different spectroscopic surveys conducted with JWST in the first five years of operations. The main contributors to this sample are JADES \citep{Eisenstein_26}, MoM (Oesch et al., \textit{in prep.}), CAPERS (Dickinson et al., \textit{in prep.}), RUBIES \citep{DeGraaf_rubies}, NEXUS \citep{Zhuang_26}, and UNCOVER \citep{Bezanson_24}. The number of objects taken from each survey can be found in the appendix of \citet{DeGraaff_25b}. The selection criteria followed to assemble the parent sample mirror those presented in Sect. \ref{sec:intro}, by including a compactness requirement in the F444W filter coupled with a series of cuts on the values of the optical and UV slopes to limit the sample to V-shaped objects (see \citealt{DeGraaff_25b} for a more complete description of the selection procedure).

For all the objects, we retrieved the spectroscopic data from the Dawn JWST Archive\footnote{\url{https://dawn-cph.github.io/dja/}} (DJA). Technical details on the data reduction pipeline employed by the DJA,  which is based on the \texttt{msaexp} library \citep{Brammer_23}, can be found in \citet{Heintz_25} and \citet{DeGraaf_rubies}. Since, as was outlined in the previous sections, our \texttt{Cloudy} modelling does not include the  treatment of the broadening of the emission lines and of the gas kinematics, we focus on the NIRSpec/PRISM data, which have a low resolution ($R\sim100$) but cover a broad wavelength range  ($\lambda\sim0.6-5.3 \ \mu$m). Moreover, since our goal is to test the reliability of our model in reproducing the red part of the LRD spectrum (i.e. redder than the Balmer break), we focus on sources with redshift of $z<6$ in order to cover a sufficiently large portion of the red continuum and have access to the Balmer and the Paschen series. Overall, the analysed sample consists of 95 objects.

\subsection{Fitting procedure}
\label{sec:fitting}

We fitted the quasi-star model to the LRDs in our sample through the following procedure. For each LRD: 
\begin{enumerate}
    \item We selected a quasi-star model from the grid computed with \texttt{Cloudy}. The reflected and diffused spectra were multiplied by the two free parameters, $\beta$ and $\gamma$, and summed up with the transmitted spectra. The output was then multiplied by a free positive $\alpha$ parameter to account for the discrete values of $L_{\rm BB}$ in our grid. The prior on $\alpha$ allows small shifts of the normalisation of the overall spectrum up to 1 dex (equal to the step of $L_{\rm BB}$ in our grid).
    
    \item We selected an SSP from the grid generated with FSPS. Given the normalisation of these spectra (Sect. \ref{sec:host}), we multiplied the SSP for a free parameter, $M_\ast$, related to the stellar mass of the host galaxy in the range $10^6-10^{12} \ M_\odot$.
    
    \item We accounted for possible dust attenuation in the host galaxy by applying a standard \citet{Calzetti_00} law with a free $A_{\rm v}$ parameter in the range 0-5 mag. Only the host galaxy includes dust attenuation, while no attenuation was applied to the quasi-star spectrum for consistency with the assumed primordial abundances.
    
    \item We summed up the two spectra and redshift the output to the spectroscopic redshift of the considered LRD included in the catalogue from the DJA.
    
    \item We resampled the obtained spectrum to the (wavelength-dependent) PRISM resolution with the \texttt{SpectRes} library \citep{Carnall_17}.
    
    \item The free parameters were then optimised by minimising the $\chi^2$ with the considered LRD. Given the primordial abundances employed in our \texttt{Cloudy} modelling, we did not consider in the $\chi^2$ computation any emission line other than hydrogen (see Sect. \ref{sec:He_lines} for a discussion on the helium lines).
\end{enumerate}

This procedure allowed us to obtain, for each LRD, a grid of $\chi^2$ values covering each combination of quasi-star and host galaxy. To account for the likely degeneracies among the free parameters, we employed a Bayesian approach in the estimation of the best-fitting quantities. Specifically, we assigned to each model a weight, $w_i=e^{-\chi^2/2}$, and computed the weighted average and related uncertainties as
\begin{equation}
    \bar{x}=\frac{\sum_i x_i \ w_i}{\sum_i w_i}
\end{equation}
and
\begin{equation}
\label{eq:uncertainty}
    \bar{\sigma}^2=\frac{\sum_i (x_i - \bar{x})^{2} \ w_i}{\sum_i w_i} \ .
\end{equation}
To account for the discrete values in our grid of parameters, we adopted a minimum uncertainty equal to half the steps reported in \Tab\ref{tab:params}. Since the luminosity of the black body ($L_{\rm BB}$) is the main responsible for the normalisation of the spectrum, we corrected the output values by the normalisation factor $\alpha$ multiplying the whole spectrum (see Sect. \ref{sec:fitting})

After removing about ten cases in which the modelling does not converge ($\chi^2\gg10$), we obtained a good fit for a total of 86 LRDs from the \citet{DeGraaff_25b} sample. A table containing the inferred parameters is published together with this study (Appendix \ref{sec:appendix_A}). Four examples of successful fittings are shown for reference in Fig. \ref{fig:examples}, while the rest of the sample is reported in Appendix \ref{sec:appendix_B}.

\subsection{Physical properties from spectral fitting}
\label{sec:properties_results}

\begin{figure}
    \centering
    \includegraphics[width=\linewidth]{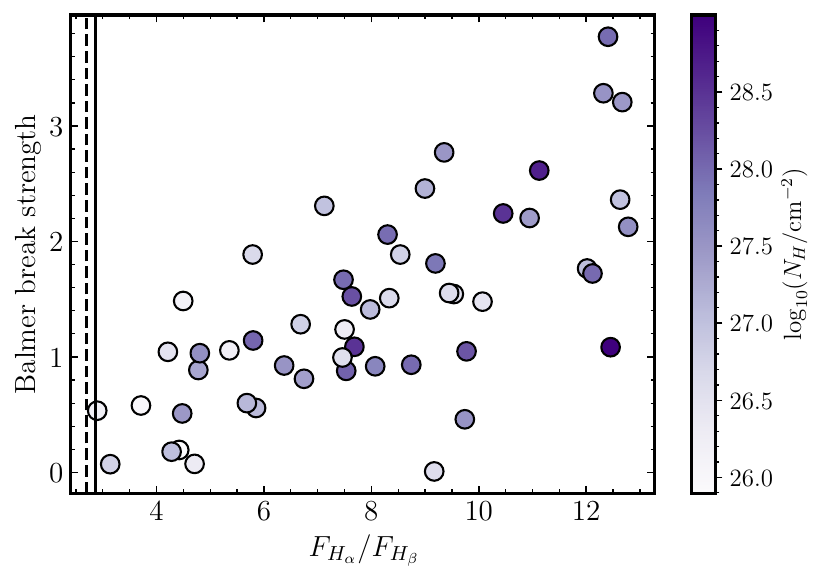}
    \caption{Empirical correlation between the strength of the Balmer break and the Balmer decrement ($F_{H_\alpha}/F_{H_\beta}$) observed by previous studies (e.g. \citealt{DeGraaff_25b,Matthee_26}) and its possible explanation in the quasi-star scenario as an effect of the increasing hydrogen column density. The values on the two axes are taken from the catalogue by \citet{DeGraaff_25b}, while the values of $N_{\rm H}$ (colour axis) are computed from the best-fitting \texttt{Cloudy} models obtained in Sect. \ref{sec:fitting}. The reference values for the Balmer decrement in case A and B recombination are reported as vertical solid and dashed black lines.}
    \label{fig:decrement}
\end{figure}

\begin{figure}
    \centering
    \includegraphics[width=\linewidth]{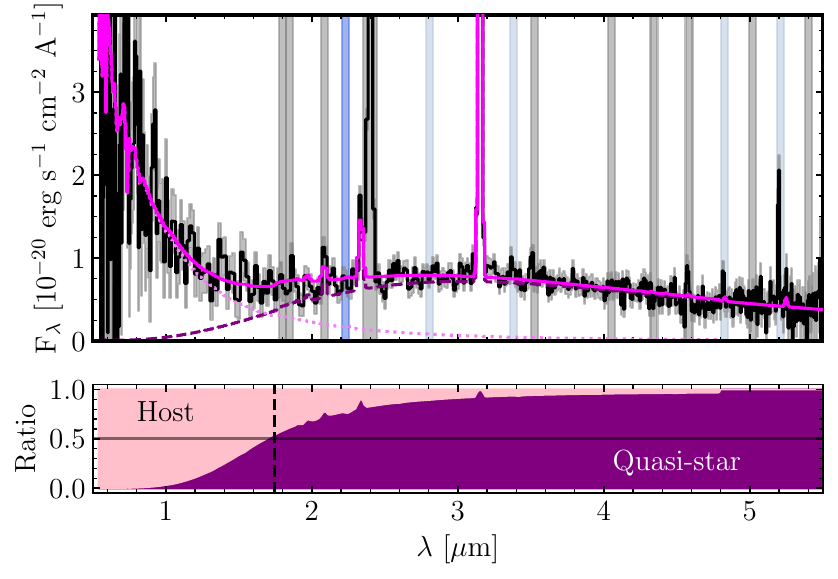}
    \caption{Example of an LRD fitted by our \texttt{Cloudy} model where the quasi-star shows a non-negligible contribution to the blue-optical spectrum bluer than the Balmer break. The LRD is ID-4820 taken from GO-2198 (PI: L. Barrufet), the colour-coding is the same as in Fig. \ref{fig:examples}, while the lower panel shows the fractional contribution of the host galaxy and of the quasi-star to the total flux of the LRD (in pink and violet, respectively). We report, for reference, the redshifted wavelength of the Balmer break (vertical dashed line) and the 50\% level (horizontal solid line).}
    \label{fig:ratio}
\end{figure}

\begin{figure*}
    \centering
    \includegraphics[width=\linewidth]{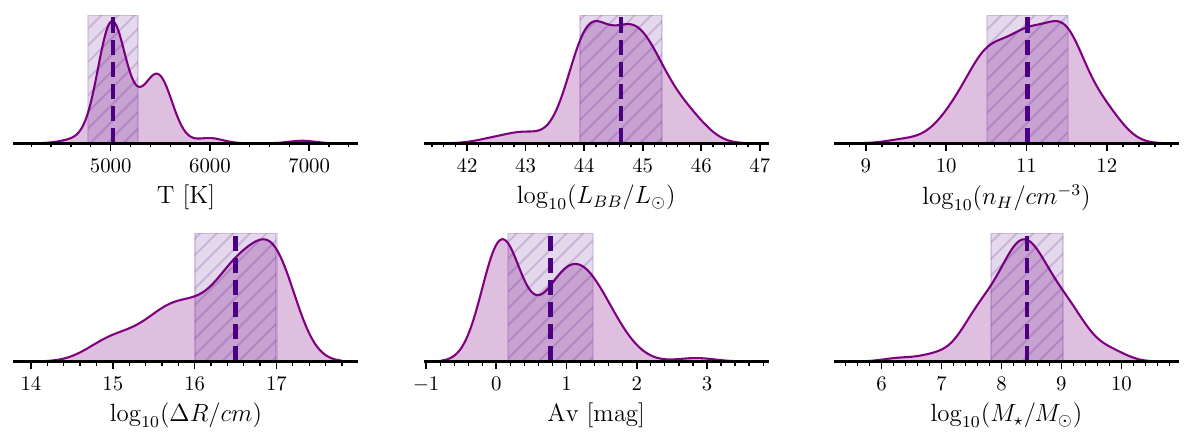}   
    \caption{{Distribution of the main parameters estimated through the fitting procedure presented in Sect. \ref{sec:fitting} applied to the LRDs in the sample by \citet{DeGraaff_25b}. The dashed vertical line indicates the median, while the shaded area report the median uncertainties computed following Eq. \ref{eq:uncertainty}.}}
    \label{fig:histograms}
\end{figure*}

As shown in Fig. \ref{fig:examples} and Appendix \ref{sec:appendix_B}, our model correctly reproduces the majority of the LRDs in the sample by \citet{DeGraaff_25b}. From a quantitative point of view, the median reduced $\chi^2$ (once the emission lines other than hydrogen are masked and the few objects with a failed modelling are removed from the sample) is of the order of 3. Small deviations can be seen in the shape of the UV continuum and in the strength of the emission lines. A possible cause of this can be found in the limited range of parameters employed in the description of the quasi-star (Sect. \ref{sec:model} and Tab. \ref{tab:params}) and of the SSP of the host galaxy (Sect. \ref{sec:host}). Moreover, our setup does not include any nebular emission from the host galaxy, which could contribute to the emission lines observed in the composite spectrum. Such contributions could be in principle unveiled by separating the broad and narrow components of the emission lines with higher-resolution data (see e.g. \citealt{Matthee_26}).

The best-fitting parameters obtained by fitting our models to the objects in the sample by \citet{DeGraaff_25b} through the procedure presented in Sect. \ref{sec:fitting} are {shown in Fig. \ref{fig:histograms}. There, we also report the median values and the median uncertainty computed following Eq. \ref{eq:uncertainty}. his last quantity is useful to understand the degree of degeneracy affecting each parameter.}

Overall, our analysis points towards a source of energy emitting a bolometric luminosity of $L_{\rm BB}\sim10^{44.4} \ {\rm erg \ s}^{-1}$ and surrounded by a zone of saturated convection, overall emitting as a black body with the same normalisation and with a median temperature of $T_{\rm BB}\sim5000 \ {\rm K}$. By applying the Stefan-Boltzmann law (Eq. \ref{eq:boltzmann}), we can derive an approximate physical scale for this convective envelope, being of the order of $R_{\rm conv}\sim10^{16.4} \ {\rm cm}$, that is of the order of 1500 AU. Such a convective zone is then surrounded by a scattering layer of dense gas with $n_{\rm H}\sim10^{11} \ {\rm cm}^{-3}$ and a radius of $\Delta R\sim10^{16.2} \ {\rm cm}$, that is approximately a thousand astronomical units. As was already noted in Sect. \ref{sec:properties}, the main parameter responsible for the spectral properties of our quasi-stars is the hydrogen column density. As shown in Fig. \ref{fig:decrement}, the increasing values of $N_{\rm H}$ correlate with the strength of the Balmer break (\textit{p} value $10^{-2}$) and with the value of the Balmer decrement (\textit{p} value $10^{-4}$), as expected in a collisionally ionised gas deviating from the standard case B recombination (see e.g. \citealt{Adams_74}). This property allows us to explain the empirical correlation observed between these two quantities observed by previous studies (e.g. \citealt{DeGraaff_25b,Matthee_26}).

Since the output spectrum depends on the integrated value of the column density, the radius of the scattering layer is degenerate with the assumed density profile. The values {shown in Fig. \ref{fig:histograms}} can therefore be considered as a lower limits, since any non-uniform profile would necessarily need a longer physical scale to obtain the same integrated column density. 

The spectrum produced by the quasi-star is then reprocessed by a diffuse medium, whose effective number of clouds (derived from the median values of the two $\beta$ and $\gamma$ parameters) is of the order of a few hundred. We emphasise, however, that the number of clouds, as well as the geometrical setup, is strongly degenerate with the physical parameters of the single clouds. Hence, it should be considered only when coupled with the parameters presented in Sect. \ref{sec:model}.

The whole quasi-star system is then hosted in a galaxy, providing the rest-UV emission of the spectrum and the dust attenuation. As visible in Fig. \ref{fig:ratio}, in some cases the quasi-star can be responsible for a non-negligible fraction of the blue-optical spectrum at wavelengths bluer than the Balmer break (even though the UV radiation in our fits is always dominated by the host galaxy emission). {In the (strong) assumptions presented in Sect. \ref{sec:host} about the stellar ages and metallicity of the SSPs employed in our modelling,} the hosting galaxies can be approximated as low-mass stellar populations ($M_\ast\sim10^{8.2} \ M_\odot$) and with little dust attenuation $A_{\rm V}\sim0.6 \ {\rm mag}$. {These values are particularly uncertain since the quasi-star emission dominates the rest-NIR part of the spectrum (giving the most stringent constraints on the stellar masses); therefore, they should be interpreted as indicative.} In any case, the obtained values of $M_\ast$ and $A_{\rm V}$ are broadly consistent with constraints reported in other studies based on galaxy clustering (e.g. \citealt{Matthee_25,Pizzati_25,Lin_26c}), metallicity (e.g. \citealt{Ivey_26}), and photometry at longer wavelengths (e.g. \citealt{Casey_24,Chen_25b}).

It is also interesting to notice that -- in a minority of cases {(see e.g. the spectra of CAPERS-EGS-19300 or JADES-GDS-2993 in the appendix)} -- the Balmer break observed in LRDs is found to come from the host galaxy (especially in cases with an older stellar population and higher stellar mass). This result is in agreement with the recent study by \citet{Cloonan_26} highlighting how a sub-class of LRDs presents a resolved morphology at the Balmer break wavelength. Similarly, a larger contribution from the host galaxy can also explain the low fraction of LRDs where the turning point of the V-shaped SED is not located exactly at the wavelength of the Balmer break (see e.g. \citealt{Setton_25,Barro_25}).

\subsection{Estimating the black hole mass}
\label{sec:BH_Masses}

One of the main features of the model presented in this paper concerns the origin of the emission lines observed in LRD spectra. While in classical AGN these lines are generated by the interaction of the ionising radiation emitted by the accreting SMBH with the clouds in the BLR, in quasi-stars they are collisionally generated within the dense gas layer surrounding the convective zone. The main observational consequence of this property is that we cannot reconstruct the properties of the radiation field at the centre of the quasi-star. However, the geometrical considerations performed in Sect. \ref{sec:properties_results} tell us that the source of energy must emit a quite large luminosity ($L_{\rm BB}\sim10^{44.4} \ {\rm erg \ s}^{-1}$) in a limited physical scale ($\Delta R \sim 1500 \ {\rm AU}$), clearly pointing towards an accreting SMBH as the source of the radiation field. 

In such a scenario, however, it is challenging to estimate the mass of the SMBH. In single-epoch measurements, the broadening of the emission lines is assumed to be due to the gravitational potential of the black hole (see e.g. \citealt{Greene_05}), which is not true in quasi-stars. At the same time, differently from the BH* model by \citet{Naidu_25}, the mass of the SMBH (and in general the properties of the accretion disk) does not enter as a free parameter in our model, since the thermalisation of the SMBH radiation in the saturated convection zone produces a BB spectrum not including any information on the original emission with the only exception of the integrated luminosity. As a consequence, the masses of the SMBH cannot be directly measured from the observed spectrum. 

\begin{figure}
    \centering
    \includegraphics[width=0.49 \linewidth]{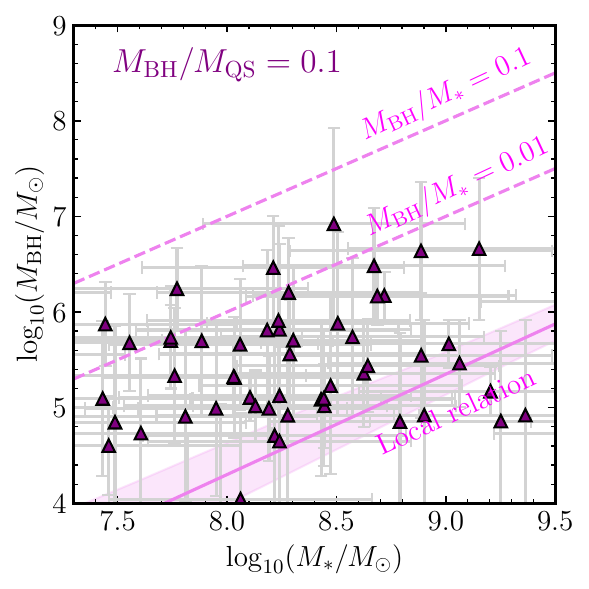} \  \includegraphics[width=0.49 \linewidth]{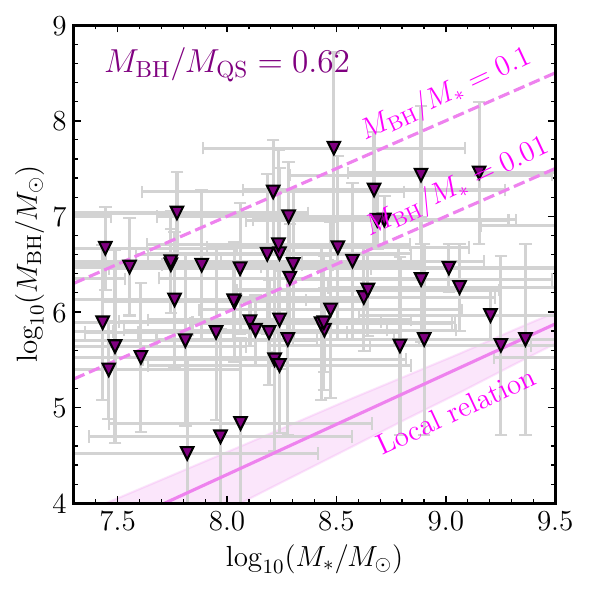}
    \caption{Comparison between the black hole masses and the stellar masses of the host galaxies for the LRDs analysed in Sect. \ref{sec:results}. The black hole masses are estimated from the bolometric luminosities of the quasi-star by assuming a ratio of $M_{\rm BH}/M_{\rm QS}$ \citep{Coughlin_24}. Two extreme values of $M_{\rm BH}/M_{\rm QS}=0.1$ (minimum for having a LRD-like spectrum) and $M_{\rm BH}/M_{\rm QS}=0.62$ (maximum to have a stable quasi-star) are reported in the two panels. The stellar masses were derived as the normalisation factor of the SSPs employed in the fitting (Sect. \ref{sec:host}). The solid magenta line and the shaded area report, respectively, the local relation by \citet{Raines_15} and its intrinsic scatter.}
    \label{fig:BHMasses}
\end{figure}

A possible solution relies on the physical model of the quasi-star. According to the theoretical studies focusing on these objects \citep{Begelman_08,Coughlin_24,Begelman_26}, the overall quasi-star emits nearly at its Eddington limit. Therefore, we can relate its luminosity ($L_{\rm BB}$ in our model) to its mass through the classical relation
\begin{equation}   
L_{\rm BB}=L_{\rm Edd}=1.2\times10^{38} \  \left(\frac{M_{\rm QS}}{M_\odot}\right) \ {\rm erg \ s}^{-1} \ ,
\end{equation}
where with $M_{\rm QS}$ we indicate the total mass of the quasi-star. For the median luminosity {shown in Fig. \ref{fig:histograms}}, this corresponds to a median mass of $M_{\rm QS}\sim2\times10^6 \ M_\odot$. This mass can then be linked to that of the SMBH through stability arguments. The above mentioned studies found that the saturated convection zone is present in quasi-stars with $M_{\rm BH}/M_{\rm QS}>0.1$. Since in our model such a zone is responsible for the overall form of the LRD spectrum, we can assume this ratio as a lower limit on the actual value of $M_{\rm BH}/M_{\rm QS}$. At the same time, the same studies found that the SMBH at the centre of the quasi-star can accrete up to $M_{\rm BH}/M_{\rm QS}\sim0.62$, after which the overall object becomes unstable \citep{Coughlin_24}. Hence, this value translates into an upper limit on $M_{\rm BH}$. The median black hole masses obtained in these two limiting cases are, respectively, $M_{\rm BH}\sim10^{5.3} \ M_\odot$ and $M_{\rm BH}\sim10^{6.1} \ M_\odot$, while the whole distributions in the two cases are shown in Fig. \ref{fig:BHMasses}. It is interesting to notice that both values translate into a super-Eddington accretion for the SMBH at the centre of quasi-stars (spanning from $\lambda_{\rm Edd}=L/L_{\rm edd}\sim10$ to $\lambda_{\rm Edd}\sim1.5$). This result is in agreement with the findings by other studies suggesting similar accretion rates (see e.g. \citealt{Lambrides_24,Secunda_26,Liu_26,Liu_26b}). 

Figure \ref{fig:BHMasses} also shows the comparison between the estimated black hole masses and the orientative stellar masses of the host galaxies estimated in Sect. \ref{sec:properties_results}. It is possible to notice how the upper limit on $M_{\rm BH}/M_{\rm QS}$ produces black hole masses in tension with the common values of $M_{\rm BH}/M_\ast$ observed in the local Universe (e.g. \citealt{Raines_15}), even though at least one order of magnitude lower than what is commonly obtained by applying standard dynamical measurements to LRDs (see e.g. \citealt{Jones_25,Maiolino_25,Ivey_26}). This tension becomes less significant (but still present) when the lower limit values are assumed (left panel of Fig. \ref{fig:BHMasses}). We also underline that the estimated stellar masses for the host galaxies are obtained in the assumption of $Z=0.1 \ Z_\odot$: recent observations (e.g. \citealt{Ivey_26}) suggest that the host galaxies of LRDs could be significantly more metal-poor (up to $Z\sim0.02 \ Z_\odot$). We verified that by assuming a lower metallicity of $Z=0.05 \ Z_\odot$ would produce an increase of $\sim0.5 \ {\rm dex}$ in stellar mass without significantly affecting the quality of the fits and further reducing the tension with the local relation by \citet{Raines_15}. {However, we further highlight how the estimation of the stellar masses for the host galaxies presented in the previous sections is uncertain: a symptom of that is the large scatter on the $M_{\rm BH}/M_\ast$ relation, significantly larger than the observed scatter on the $M_{\rm UV}/M_{\rm opt}$ relation (see e.g. \citealt{DeGraaff_25b}). A more accurate determination of the host masses would require a subtraction of the quasi-star model and -- most likely -- higher-resolution data to disentangle the line emission of the central engine from that of the host galaxy.}

\subsection{Other observables of LRDs}

As was shown in the previous sections, our \texttt{Cloudy} modelling of the quasi-star simply reproduces the strength of the hydrogen emission lines, the shape of the optical-NIR continuum of LRDs, and -- with the additional contribution of a star-forming host galaxy -- of the UV continuum. Two observables of LRDs that are not natively reproduced by our model (and that therefore need an additional component) are the helium lines and the possible presence of hot dust.

\subsubsection{Helium emission}
\label{sec:He_lines}

Relatively bright emission lines of neutral helium (\ion{He}{I} $\lambda5876$, $\lambda7065$, $\lambda10028$, and $\lambda10830$) are commonly observed in our sample (nearly 75\% of the LRDs whose spectra are reported in Appendix \ref{sec:appendix_B}). On the contrary, emission lines from ionised helium (\ion{He}{II}; e.g. $\lambda4686$) are rarely observed (less than 10\% of the sample). The resolution of the PRISM is not adequate to measure the width of these lines in the spectra considered here and -- therefore -- establish if they are narrow or broad. However, other studies based on higher-resolution data of LRDs reported that neutral helium lines are observed to be broad (see e.g. \citealt{Torralba_26}), suggesting that they could be produced somewhere in the engine of LRDs and not in the host galaxy.

Helium emission lines are not expected in any of our \texttt{Cloudy} models, since helium cannot be either photoionised by the relatively cold BB radiation that we infer in our fits or collisionally excited at the density and temperature of the gaseous shell, due to its higher ionisation potential compared to hydrogen.Emission lines of \ion{He}{I} and \ion{He}{II} are commonly observed in AGN, where the harder radiation field can photoionise helium (e.g. \citealt{Almog_89,Riffel_06}), and would be therefore generated in proximity of the accreting SMBH at the centre of the quasi-star. However, these emission lines are not expected to reach the surface of the object, since the ionising radiation emitted from the accreting SMBH is fully thermalised inside the convective core. At the same time, the rarer occurrence of HeII emission lines in LRDs can be interpreted as evidence that photoionisation from the accreting SMBH is not the only mechanism to explain helium emission lines.

One possible way in which quasi-stars could produce helium emission lines is through the presence of a hot coronal gas surrounding the scattering layer heated up by the presence of magnetic fields, analogously to what is commonly observed in the coronal gas surrounding stars. Such a mechanism would be analogous to the solar corona, where the \ion{He}{I}$\lambda5876$ and $\lambda 10830$ emission lines are comparatively strong (see e.g. \citealt{Gabriel_95,Moses_20}). The temperature of the coronal gas (in the order of a few millions in the Sun) would make photoionisation and radiative and collisional excitation of helium possible. As recently noticed by \citet{Takasao_26}, the existence of magnetic fields is a direct consequence of the assumption about the convective core of the quasi-star, where dynamo mechanisms are expected to produce them. If further studies of the frequency and intensity ratios of HeI and HeII emission from LRDs for which the quasi-star is a good model of the optical SED are consistent with coronal emission, then these lines can be used as tracers and diagnostics of the properties of the coronal gas that surrounds the source and of its magnetic fields.

\subsubsection{Hot dust emission}
\label{sec:dust}

\begin{figure}
    \centering
    \includegraphics[width=\linewidth]{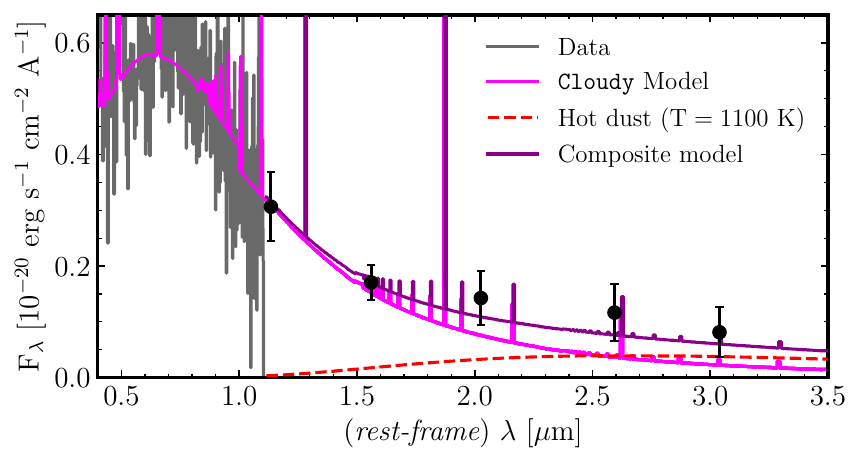}
    \caption{Example of a LRD (JADES-GDS 13329) with mid-infrared excess at MIRI wavelengths not explainable with the simple quasi-star model but likely requiring an additional component of hot dust with $T=1100 \ {\rm K}$. We report the original \texttt{Cloudy} model in magenta, the additional hot dust component in dashed red, and the composite model in violet. The photometric points are taken from the SMILES survey \citep{Rieke_24,Alberts_24}.}
    \label{fig:hot_dust}
\end{figure}

A second observable of LRDs that cannot be directly explained by the quasi-star model, at least in its current form, is the possible presence of emission from hot dust. By looking at the predictions of the model at observed wavelengths above 5.3 $\mu{\rm m}$ (i.e. outside the range covered by PRISM observations), we can compare them with MIRI photometry, when available (see an example in \ref{fig:hot_dust}). Since this information was not employed in the fitting procedure, this test represents an independent validation of our model. The model correctly meets the MIRI constraints up to rest-frame $\lambda\sim1.5 \ \mu{\rm m}$, where the photometry follows the shape of a modified BB, as shown by \citet{DeGraaff_25b}. At longer wavelengths, however, an excess flux is visible.

Currently, the evidence of this excess flux is mostly based on stacked MIRI images (e.g. \citealt{Delvecchio_25,PerezGonzalez_26,Barro_25}) and on the individual photometry of a small number of single objects, such as JADES-GDS 13329 (shown in Fig. \ref{fig:hot_dust}). For these reasons, it is still an open question if the flux excess is a feature of the whole population of LRDs or if it appears only in a (potentially substantial) subset of the sources. In both cases, this mid-infrared excess of LRDs has been recently interpreted as evidence of hot dust ($T\sim10^3 \ {\rm K}$; e.g. \citealt{Delvecchio_25,PerezGonzalez_26,Madau_26}). As a test, we indeed verified that the MIRI photometry of JADES-GDS 13329 is consistent with the presence of an additional BB component with $T\sim1100 \ {\rm K}$ (Fig.  \ref{fig:hot_dust}). 

The best-fit spectra of our quasi-star model have the shape of a modified BB and thus cannot, at least in their present form, explain the excess at mid-infrared wavelengths. If such excess emission is actually generated by hot dust, then this dust should be located in the inner regions of the quasi-star, in the proximity of the accreting SMBH that provides the source of the heating radiation. This location of the hot dust is a direct consequence of the lack of dust attenuation in the scattering layer and in the diffuse clouds inferred by our fits, and of the fact that the temperature of the convective zone ($T_{\rm BB}\sim5000 \ K$) is above the sublimation point of dust ($T\sim1500 \ {\rm K}$; e.g. \citealt{Barvainis_87}). In this case, this dust could have formed from the metals generated through nucleosynthesis in the previous evolutionary stages of the very massive star that originated the quasi-star. To survive in the innermost regions of the quasi-star, however, the dust should be somehow shielded from the direct radiation of the accretion disk (e.g. by dense gas) and and also be sufficiently distant from the base of the convective layer. How the thermal radiation from the hot dust, emitted at rest-frame mid-infrared wavelengths, can escape the quasi-star and be observed is also a point that needs to be further understood. In the model of the quasi-star, it is assumed that the convective layer is completely optically thick, (i.e. with very high optical depth at all wavelengths). This assumption is clearly an over-simplification, since in normal astrophysical systems the optical depth is found to decrease with increasing wavelength. In this hypothesis, the hot dust radiation could escape relatively unprocessed the inner structure of the quasi-star and reach the surface. If future studies can validate such a hypothetical scenario, then the hot dust emission observed in the SED of LRDs could be used as a powerful probe of their internal structure. 

\section{Discussion}
\label{sec:discussion}

In the previous sections, we showed how a relatively simple model of a quasi-star can reproduce most of the main observables linked to LRDs in a physically consistent way. Several studies so far have pointed out that LRDs are unlikely to represent a single and homogeneous population of objects, with several properties strongly varying from LRD to LRD. Some of them include the possible presence of an extreme Balmer breaks (suggesting larger gas reservoirs surrounding the SMBH or a lower contribution from the host galaxy; see e.g. \citealt{DeGraaff_25a,Naidu_25,Taylor_25,Ji_25,Barro_25,Billand_26}), the presence of highly ionised emission lines in the UV part of the spectrum (suggesting a direct view of the accretion disk of the SBMH; see e.g. \citealt{Akins_25b,PerezGonzalez_26,Ji_26}), and the presence of an extended components visible in the UV bands (suggesting a significant contribution from the host galaxy; see e.g. \citealt{Rinaldi_25,Ma_26a,Cloonan_26}). Therefore, the main goal of this paper is not to propose a model able to explain all LRDs, but rather show that most possible explanations are degenerate with other physical models.

\subsection{Comparison with other quasi-star models}
\label{sec:other_models}

The analysis carried out in this paper represents a first ‘empirical’ validation of the model presented by \citet{Begelman_08} and \citet{Begelman_26} explaining LRDs as late-stage quasi-stars. The best-fitting values presented in Sect. \ref{sec:properties_results} and {shown in Fig. \ref{fig:histograms}} well match the forecasts provided in \citet{Begelman_26}. In that model, LRDs represent an intermediate stage in the evolution of a massive star. In more detail, \citet{Begelman_26} consider a massive star with a total mass of the order of {$M_{\rm QS}\sim10^{4-5} \ M_\odot$} and a luminosity of $L_{\rm QS}\sim10^{42-43} \ {\rm erg \ s}^{-1}$. The subsequent evolution of the massive star produces the increase in the total mass and luminosity of nearly 1 order of magnitude, the collapse of the central region, and the consequent formation of a black hole (the ‘early quasi-star’). The black hole starts to accrete at a super-Eddington rate, leading to the formation of the saturated convection zone and the late-stage quasi-star that we model in this paper. According to the model by \citet{Begelman_26}, at this point the quasi-star has reached a total mass of $M_{\rm QS}\sim10^{6-7} \ {M_\odot}$ and a luminosity of $L_{\rm QS}\sim10^{44-45} {\rm erg \ s}^{-1}$, while the inner black hole reaches a mass larger than 10\% of the total. These quantities match well the estimates discussed in Sect. \ref{sec:properties_results} and {shown in Fig. \ref{fig:histograms}}. Overall, our analysis supports the theoretical results presented in \citet{Begelman_26} suggesting that late-stage quasi-stars can explain the observed properties of some LRDs. {A notable difference -- however -- concerns the temperature of the quasi-star. In \citet{Begelman_26}, the authors describe how these objects should have a quite narrow range of temperature, despite having a broad range of masses and luminosities. While this prediction is confirmed by our results (the vast majority of the analysed LRDs have temperatures in the range $5000-5500 \ {\rm K}$ while the luminosities span three orders of magnitude; see an analogous result presented in \citealt{DeGraaff_25b}), our modelling suggests a lower temperature than the $6000 \ {\rm K}$ forecasted by \citet{Begelman_26}. In the framework presented by the authors, this result could indicate a low convection efficiency in the optically thick envelope.}

When looking at other models present in the current literature, the quasi-star by \citet{Begelman_08} is conceptually similar to the BH* presented by \citet{Naidu_25} and \citet{DeGraaff_25b}. In the first of these models, an accreting SMBH is surrounded by a dense atmosphere of gas: the radiation from the AGN does not reach the complete thermal equilibrium with the surrounding medium, allowing the output spectrum to keep the information about the original source of energy (the mass of the black hole and the temperature of the accretion disk are, indeed, free parameters of the fitting performed in \citealt{Naidu_25}). In such a scenario, lines can be formed thanks to the ionising radiation produced by the AGN; therefore, the involved column densities are lower than those obtained in our study. Since the overall gas density of the two models are in the same order of magnitude ($n_H\sim10^{11} \ {\rm cm}^{-3}$), the black hole star by \citet{Naidu_25} is significantly smaller than our quasi-star (about 50 AU instead of the 1000 AU reported in Sect. \ref{sec:properties_results}). A possible important issue with this model resides in the stability of such an object: to explain the observed luminosity of LRDs, the black hole star should accrete in a strongly super-Eddington way ($\lambda_{\rm Edd}\sim5-10$; \citealt{Naidu_25}), which according to \citet{Begelman_08} could happen on a long timescale ($\sim500 \ {\rm Myr}$) only in the presence of an optically thick zone of saturated convection. 

This zone is included in the model empirically derived by \citet{DeGraaff_25b}, where the radiation is indeed thermalised and the output spectrum assumes the shape of a modified black body (matching the observed continuum of most LRDs). This updated model, however, does not discuss the physical origin of this spectral shape (i.e. why a modified BB and not the classic Planck function) or the origin of the spectral features such as the emission lines and the Balmer break. The model presented in this paper solves these issues. Firstly, the reprocessing of the BB radiation in the scattering layer is at the origin of the modified shape of the output spectrum, matching the observational result by \citet{DeGraaff_25b}. Secondly, our model shows that a partial ionisation can be reached in the outskirts of a quasi-star thanks to the high gas density, even in the presence of a low-temperature BB ($T_{\rm BB}\sim5000 \ {\rm K}$), allowing for the formation of emission lines. At the same time, the over-population of the low-energy levels of the hydrogen atoms are responsible for the formation of the strong Balmer breaks.

\subsection{Limitation of the model and caveats}
\label{sec:caveats}

The model discussed in this paper clearly presents some limitations when applied to the overall population of LRDs. We already discussed the possible issues with the formation of helium lines and with the possible presence of hot dust emission. Even if the addition of additional components (namely, a corona of hot gas and a dusty accretion disk in the centre of the quasi-star) could potentially solve these issues, their theoretical feasibility should be studied in more detail. In addition, some studies reported single detections of LRDs showing high-ionisation lines in their spectra (e.g. \citealt{Akins_25b,Torralba_26,Deugenio_25,Tang_25,Tang_26}). In the quasi-star scenario, such lines cannot be produced by the relatively cold BB emission and -- if generated by the accreting SMBH -- are not expected to reach the surface due to the complete thermalisation of the ionising spectrum. These properties make the quasi-star model not easily applicable to the examples of LRDs where such lines have been detected.

Together with these issues, a more fundamental point needs to be addressed. Even though the vast majority of LRDs are located at high redshifts ($z>4$; see e.g. \citealt{Kocevski_23}), several examples of low-redshifts LRDs have been presented so far (e.g. \citealt{Bisigello_25,Lin_25,Lin_26,Park_26,Chen_26}), with some examples identified up to $z\sim1.7$ \citep{Torralba_26}. The quasi-star model presented in \citet{Begelman_08,Begelman_26} relies on the collapse of a massive star with $M_\star\sim10^{4-5}$ to form these objects. Such massive stars are commonly thought to arise from low-metallicity gas (see e.g. \citealt{Begelman_08,Begelman_26}) and to have a short lifetime ($\tau\sim2-5 \ {\rm Myr}$): the identification of low-redshift LRDs is somehow challenging to reconcile with these properties. On the one hand, one needs to explain how a cloud of nearly pristine gas can persist up to the low-redshift Universe without being polluted by the chemically evolved ISM of the host galaxy. On the other hand, one needs to explain why the collapse of such a star should only take place several billion years after the formation of the host galaxy. A possible explanation has been recently suggested by \citet{Baggen_26} thanks to the observational evidence that most LRDs seem to have a neighbouring low-mass and star-forming companion. In that study, the authors suggest that those blue companions could behave as sources of Lyman-Werner radiation able to prevent the cooling of molecular hydrogen clouds in the LRD hosts, allowing for the late collapse of these clouds and late formation of massive stars. These scenarios need further investigations from our community to make the quasi-star model a reasonable explanation for the LRD properties.

\subsection{Concluding speculative remarks}

One of the main results of our analysis is to show that a simple model of quasi-star can reproduce the rest-frame optical-NIR continuum and hydrogen emission lines commonly observed in LRDs. This model cannot, however, reproduce the UV continuum. Nevertheless, it is hard to imagine a scenario where a quasi-star could be observed in isolation (i.e. with no host galaxy or stellar populations associated with it), except -- perhaps -- if powered by an accreting PBH and at very high redshift, when no appreciable star formation has yet taken place in its proximity.

In all other cases, unveiling the presence of a quasi-star requires an accurate prediction of its spectrum in order to separate its emission from that of the host galaxy and study in detail the latter (see e.g. an earlier attempt in \citealt{Sun_26}). If quasi-stars can be confirmed as the source of power of some LRDs, they will almost certainly change our view on how SMBH and AGN -- as well as ultra-dense gaseous structures -- can form early on in the cosmic history. For this reason, they are conceptually interesting.

{ As per the identification of quasi-stars powered by PBHs -- if they exist at all -- it will almost certainly require pushing the search to very high redshifts, when no star-formation had taken place in or around such objects \citep{Matteri_25}, presumably at redshifts $z\gtrsim15$. In this range, both observations and models predict the number density of JWST-detectable star-forming galaxies to be substantially lower than at earlier epochs (up to several orders of magnitude; see e.g. \citealt{Castellano_25,PGonzalez_25,Somerville_25}. This search  will also require a different approach even in the identification of the candidates.} Firstly, we will most likely need to modify the selection criteria of LRDs, because the requirement of a V-shaped SED, with its blueward rising SED, also implies the requirement to have rest-frame UV emission, (i.e. relatively unobscured star formation). Thus, a PBH-powered quasi-star candidate will be an object similar to a ‘dropout’ but whose SED to the red of the ‘break’ rises more gently and gradually than a ‘Lyman break galaxy’, reaches a peak, and then decreases roughly as $\propto \nu^{-2}$. The candidate will also have hydrogen emission and absorption features in its spectrum. In other words, it will be an object similar to a Balmer-break galaxy that is still hosting star formation, but with a bluer optical-NIR SED and with no metal lines at any wavelength, with an extremely faint emission bluer of the break (consistent with the Wien region of the BB spectrum) and with no detectable presence of dust emission, both hot and cold. We have not yet started searching for such objects, even though they could already hide as contaminants of samples of extremely high-\textit{z} galaxies (e.g. \citealt{PGonzalez_25b,Gandolfi_26}).

\section{Summary}
\label{sec:summary}

In this paper, we present the first ‘empirical’ test of the quasi-star model by \citet{Begelman_08,Begelman_26} as a possible explanation for the spectral properties commonly observed in LRDs. Thanks to a new suite of radiative-transfer simulations carried out with the code \texttt{Cloudy}, we studied whether this model is able to reproduce the shape of the rest-frame UV-to-NIR continuum as well as the strength of the hydrogen emission lines commonly observed in LRDs. Our main results are the following:

\begin{enumerate}
    \item A satisfying model of a quasi-star includes a source of energy (likely a SMBH) surrounded by an optically thick zone of saturated convection. Such objects emit a BB radiation only characterised by its total luminosity ($L_{\rm BB}$) and temperature ($T_{\rm BB}$). This radiation is then reprocessed by a layer of dense gas (the scattering layer) in direct contact with the convective shell and characterised by its hydrogen density ($n_H$) and thickness ($\Delta R$). The whole system is then surrounded by a diffuse and clumpy medium, with properties that can be assumed as analogous of those measured in the BLR of a classical AGN.\\

    \item In this model, the emission lines and the strong Balmer breaks commonly observed in LRDs are produced in the scattering layer, whose ionisation properties are mostly driven by thermal collisions. The reprocessing in this layer is also responsible for the shape of the red continuum, while the wavelengths surrounding the Balmer limit are affected by the absorption and reflection taking place in the diffuse medium.\\

    \item By fitting this model to the NIRSpec/PRISM data of the LRDs in the sample by \citet{DeGraaff_25b}, we verify that quasi-stars can reproduce the vast majority of the UV-to-NIR continuum of LRDs as well as the strength of the hydrogen emission lines.\\

    \item This procedure also allows us to estimate the best-fitting values for the free parameters in the model. Overall, the quasi-star can be described as a dense ($n_{\rm H}\sim10^{11} \ {\rm cm}^{-3})$ and small ($R\sim2000 \ {\rm AU}$) envelope of gas surrounding a low-mass SMBH ($M_{\rm BH}\sim10^{5.3-6.1} \ M_\odot$) and surrounded by a proto-BLR with an effective number of clouds of the order of a few hundred.\\

    \item In the quasi-star model presented in this paper, the rest-UV radiation observed in LRDs is mostly due to the contribution of a low-mass ($M_\ast\sim10^8 \ M_\odot$) star-forming galaxy, in the assumption of a SSP with stellar age lower than 1 Gyr and low metallicity ($Z=0.1 \ Z_\odot$). Different hypotheses would clearly bring to different estimates of the stellar mass. In more detail, the assumption of a lower metallicity ($Z=0.05 \ Z_\odot$) would produce an increase of nearly 0.5 dex in the estimated stellar mass.\\

    \item The mass of the SMBH can only be estimated from the bolometric luminosity of the quasi-star, since the broadening of the emission lines is unrelated to the gravitational field of the SMBH (as commonly assumed in single-epoch measurements). The quoted value of $M_{\rm BH}\sim10^{5.3-6.1} \ M_\odot$ is estimated in the hypothesis of Eddington-limited luminosity for the quasi-star and of a ratio $M_{\rm BH}/M_\ast$ in the range of 0.1-0.62, according to the theoretical studies by \citet{Coughlin_24} and \citet{Begelman_26}.\\

    \item Within these assumptions about the mass ratio between the SMBH and the quasi-star, and about the properties of the host galaxy, we obtain that the SMBH at the centre of quasi-stars represent between 1\% and 0.1\% of the total mass of their host galaxies. This value is lower than those obtained through the line broadening and by dynamical considerations, but is still in slight tension with the local relation by \citet{Raines_15}.\\

    \item While the quasi-star model natively accounts for the shape of the continuum and the strength of the emission lines, it does not reproduce the presence of helium emission lines and the possible emission from hot dust. These additional components require modifications of the simple quasi-star model, such as the addition of a coronal gas heated by magnetic fields, the presence of dust in the inner regions of the quasi-star, and a wavelength-dependent optical depth decreasing for $\lambda\gtrsim1 \ \mu{\rm m}$.
    
\end{enumerate}

Overall, this study represents a first attempt at analysing in detail the radiative transfer in the quasi-star model. However, it is clearly limited by some of the intrinsic characteristics of the \texttt{Cloudy} model, in particular in the determination of the resolved shape of emission lines. The natural continuation of this study is the employment of more advanced codes to properly account for the non-statical kinematics of quasi-stars.

\section*{Data availability}
The catalogue of the inferred parameters for the LRDs analysed in this work is available in electronic form at the CDS via anonymous ftp to \url{cdsarc.u-strasbg.fr} (130.79.128.5) or via \url{http://cdsweb.u-strasbg.fr/cgi-bin/qcat?J/A+A/}. A description of the columns can be found in Appendix \ref{sec:appendix_A}.

\begin{acknowledgements}
We thank the anonymous referee for the constructive report and for the comments that helped us improving the initial manuscript. We thank A. de Graaff and collaborators for publicly releasing their catalogue of LRDs. FG thanks Livia Vallini for the useful discussion about the \texttt{Cloudy} modelling and for the comments on an early version of the study. The data products presented herein were retrieved from the Dawn JWST Archive (DJA). DJA is an initiative of the Cosmic Dawn Center (DAWN), which is funded by the Danish National Research Foundation under grant DNRF140.
\end{acknowledgements}

\bibliography{biblio}
\onecolumn
\begin{appendix}

\section{Catalogue header}
\label{sec:appendix_A}
\begin{table*}[h]
    \centering
    \caption{Description of the table published with this work.}
    \begin{tabular}{ccl}
        Column & Units & Description \\
        \hline
        \hline
        ID & - & ID from the DJA \\
        RA & deg [J2000] & Right ascension \\
        Dec & deg [J2000] & Declination\\
        PID & - & Programme ID of the JWST observations \\
        z\_spec & - & Spectroscopic redshift from the DJA \\
        T & K & Temperature of the BB spectrum \\
        T\_Err & K & Uncertainty on T \\
        L & erg s$^{-1}$ & Luminosity of the BB spectrum (log)\\
        L\_Err & erg s$^{-1}$ & Uncertainty on L (log) \\
        Nh & cm$^{-3}$ & Hydrogen density of the scattering layer (log)\\
        Nh\_Err & cm$^{-3}$ & Uncertainty on Nh (log)\\
        Depth & cm & Thickness of the scattering layer (log)\\
        Depth\_Err & cm & Uncertainty on Depth (log)\\
        Av & mag & Dust attenuation of the host galaxy\\
        Av\_Err & mag & Uncertainty on Av\\
        Mass & $M_\odot$ & Stellar mass of the host galaxy (log)\\
        Mass\_Err & $M_\odot$ &  Uncertainty on mass\\
        $M_{\rm BH,Upp}$  & $M_\odot$ &  Mass of the SMBH (assuming $M_{\rm BH}/M_{\rm QS}=0.62$)\\
        $M_{\rm BH,Upp}$\_Err  & $M_\odot$ & Uncertainty on $M_{\rm BH,Upp}$\\
        $M_{\rm BH,Low}$  & $M_\odot$ &  Mass of the SMBH (assuming $M_{\rm BH}/M_{\rm QS}=0.1$)\\
        $M_{\rm BH,Low}$\_Err  & $M_\odot$ & Uncertainty on $M_{\rm BH,Low}$\\
        \hline
    \end{tabular}
    \label{tab:header}
\end{table*}

\newpage
\section{Full sample}
\label{sec:appendix_B}

\begin{figure*}[h]
\centering
\includegraphics[width=0.23\linewidth]{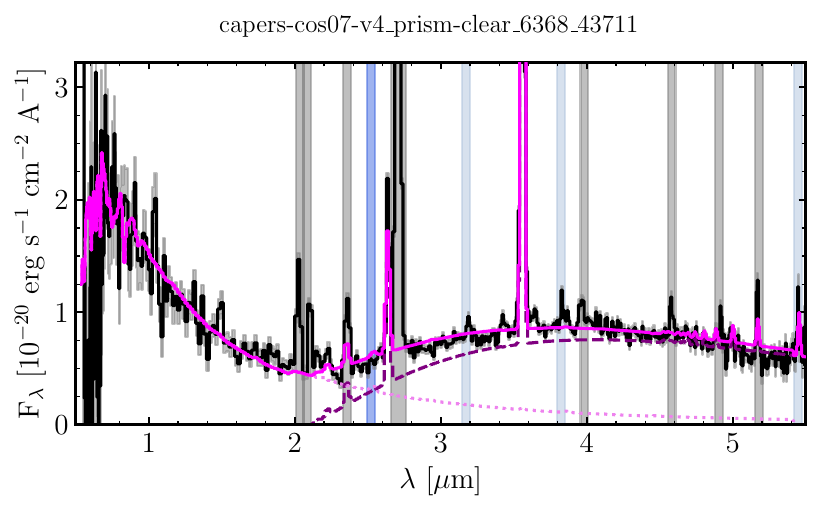}
\includegraphics[width=0.23\linewidth]{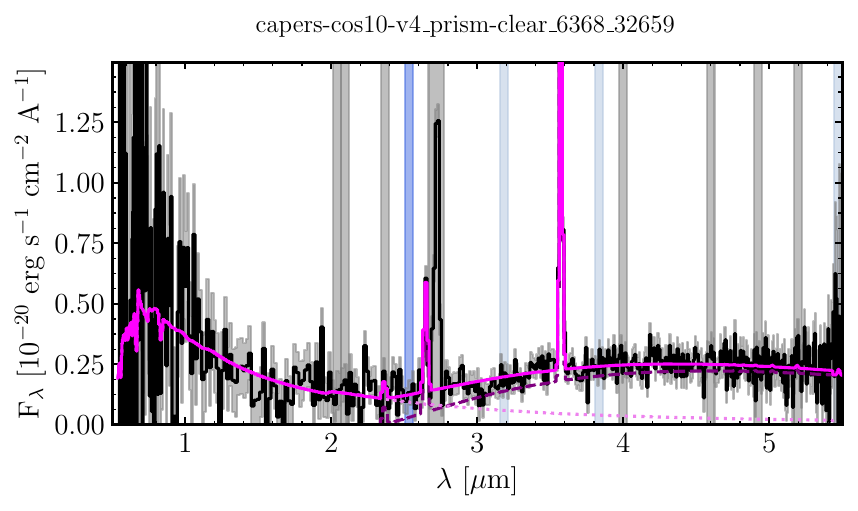}
\includegraphics[width=0.23\linewidth]{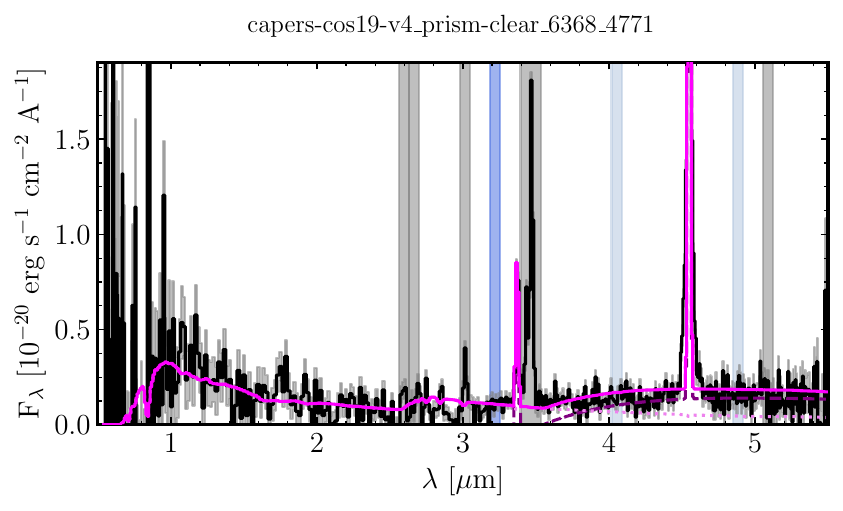}
\includegraphics[width=0.23\linewidth]{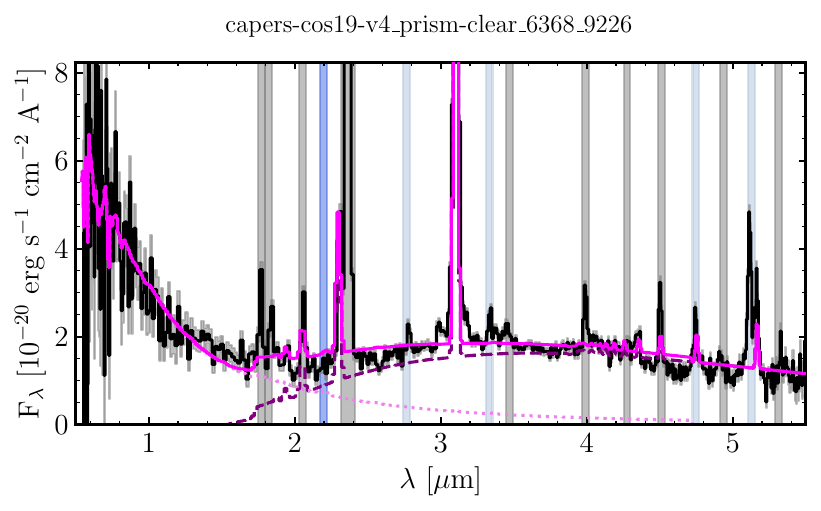}

\includegraphics[width=0.23\linewidth]{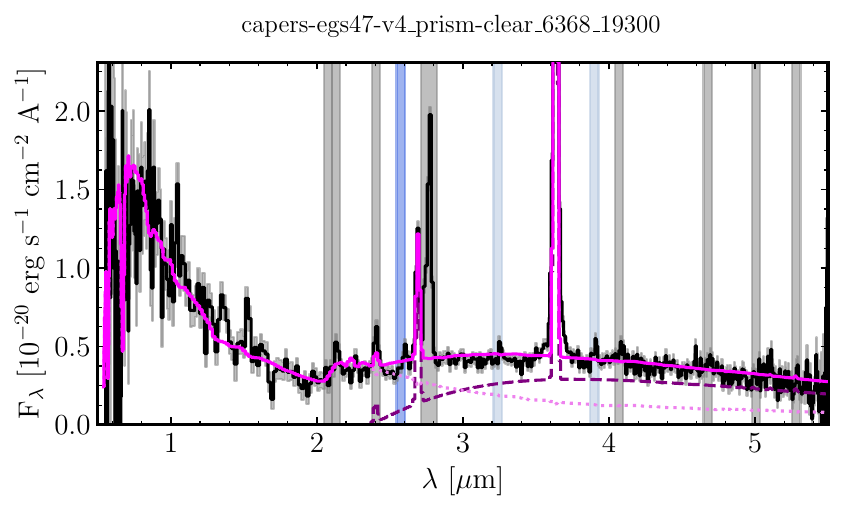}
\includegraphics[width=0.23\linewidth]{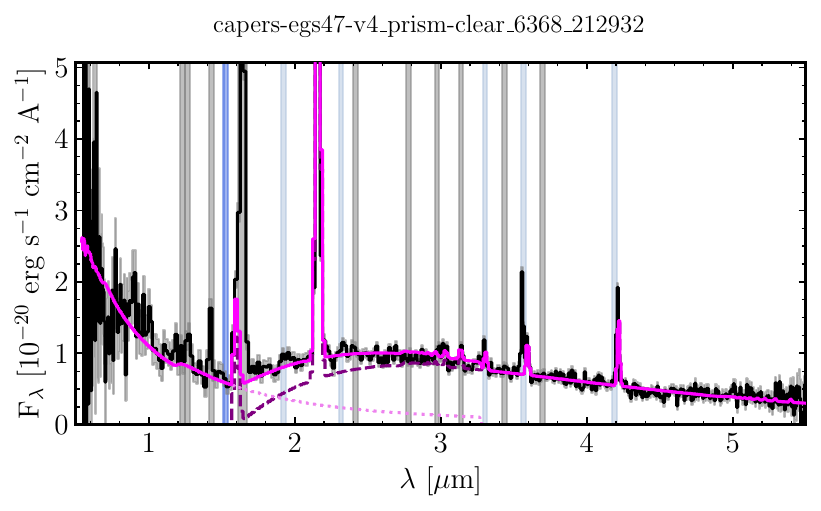}
\includegraphics[width=0.23\linewidth]{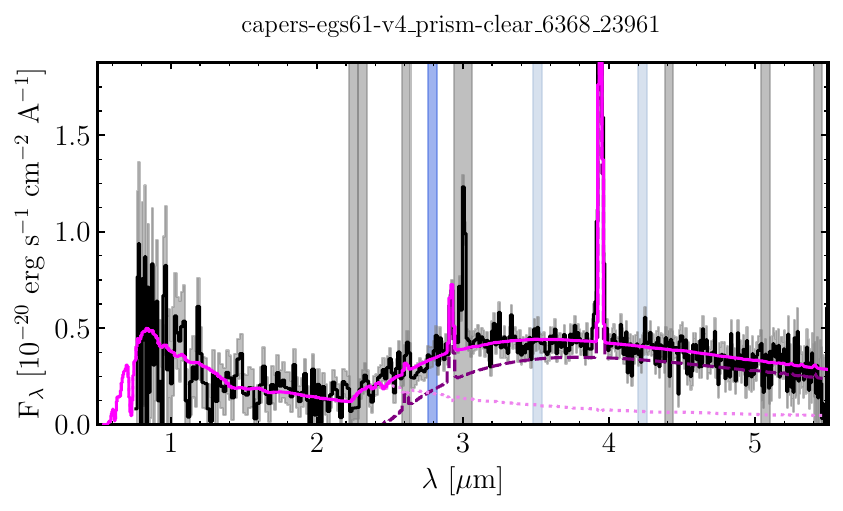}
\includegraphics[width=0.23\linewidth]{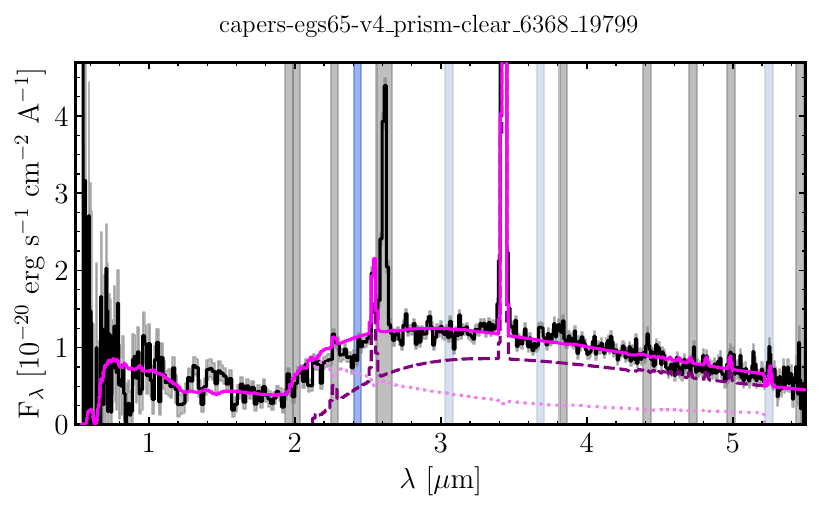}

\includegraphics[width=0.23\linewidth]{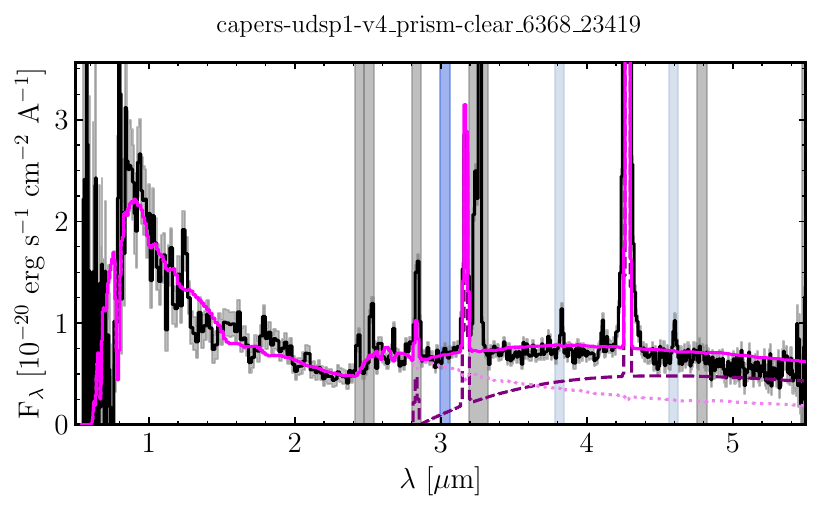}
\includegraphics[width=0.23\linewidth]{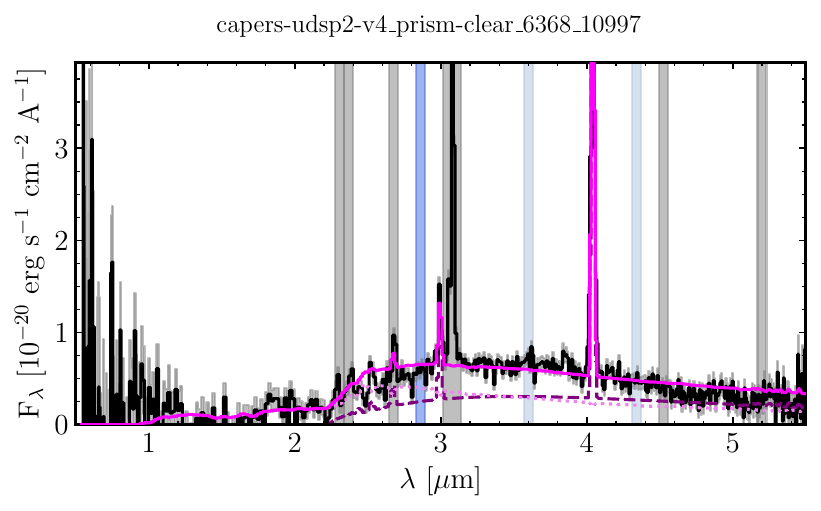}
\includegraphics[width=0.23\linewidth]{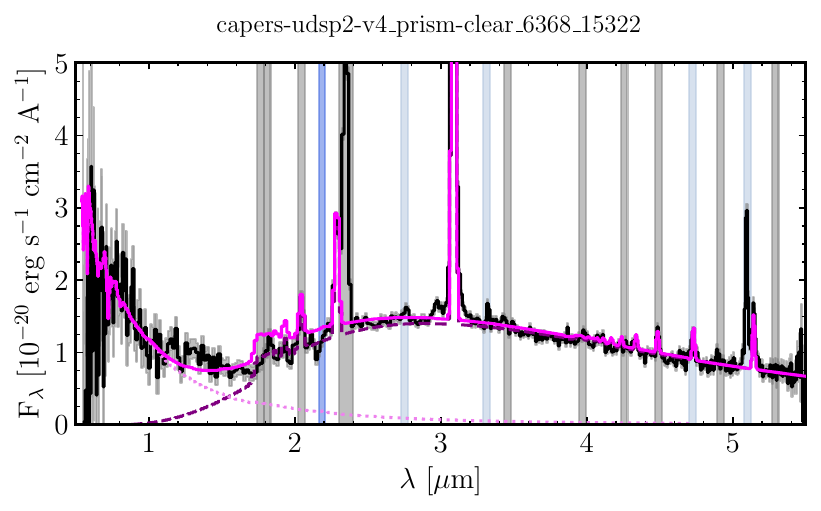}
\includegraphics[width=0.23\linewidth]{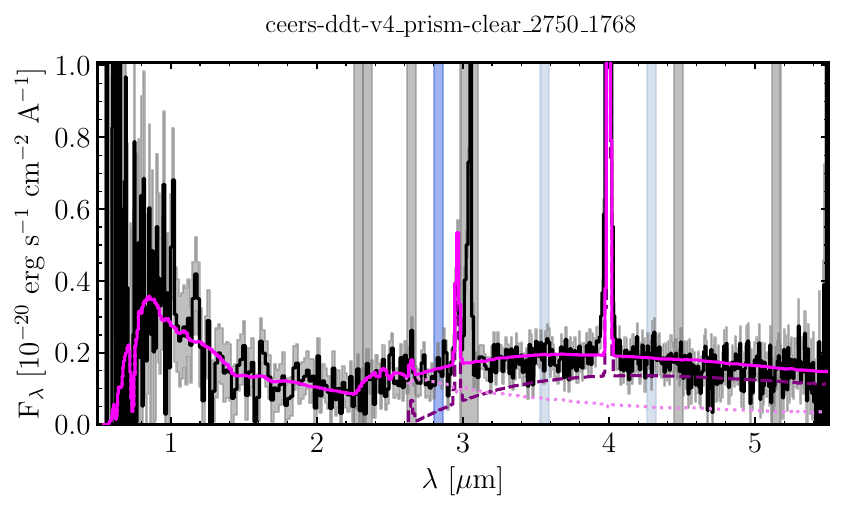}

\includegraphics[width=0.23\linewidth]{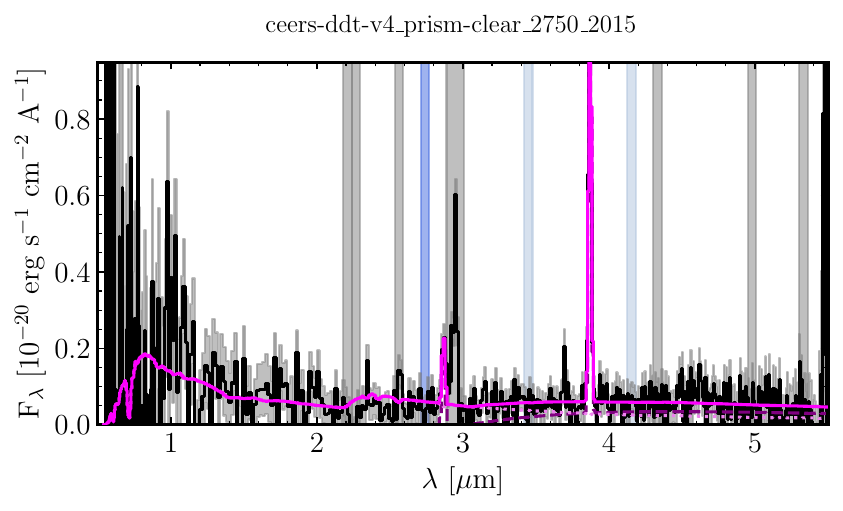}
\includegraphics[width=0.23\linewidth]{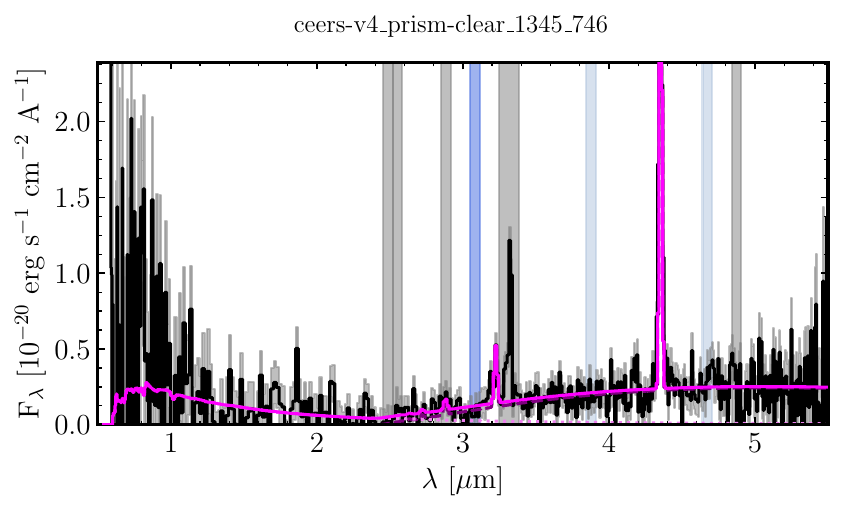}
\includegraphics[width=0.23\linewidth]{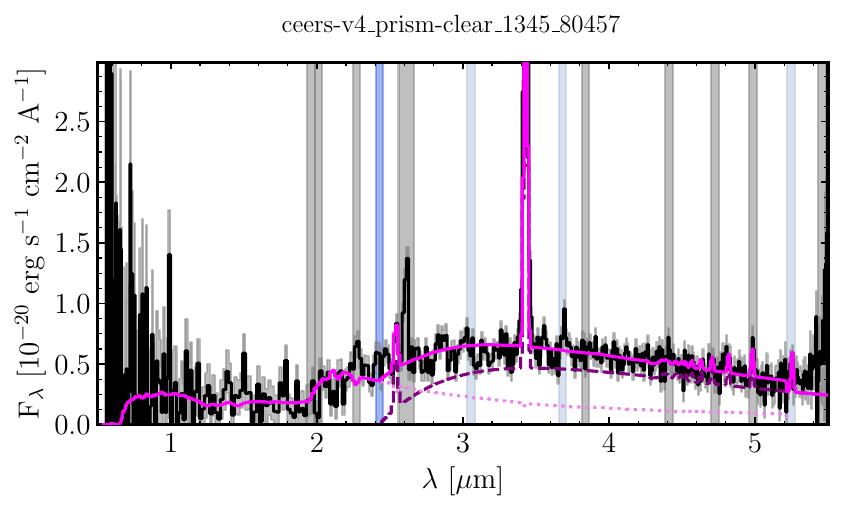}
\includegraphics[width=0.23\linewidth]{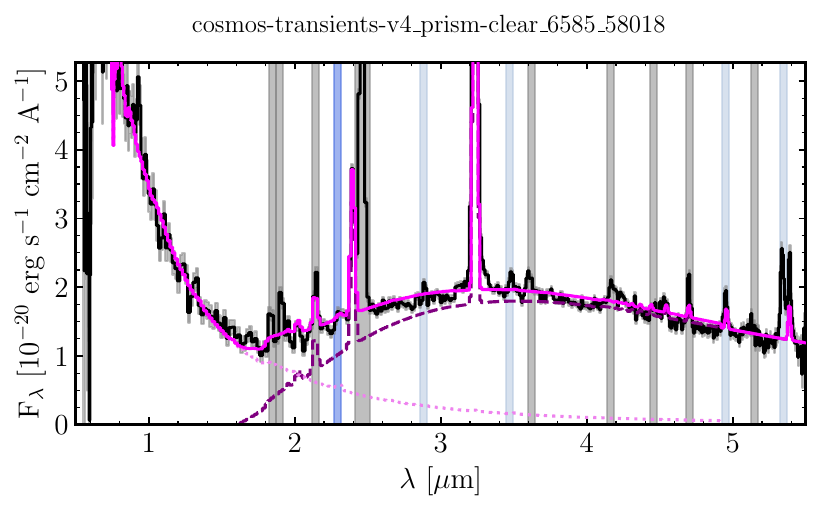}

\includegraphics[width=0.23\linewidth]{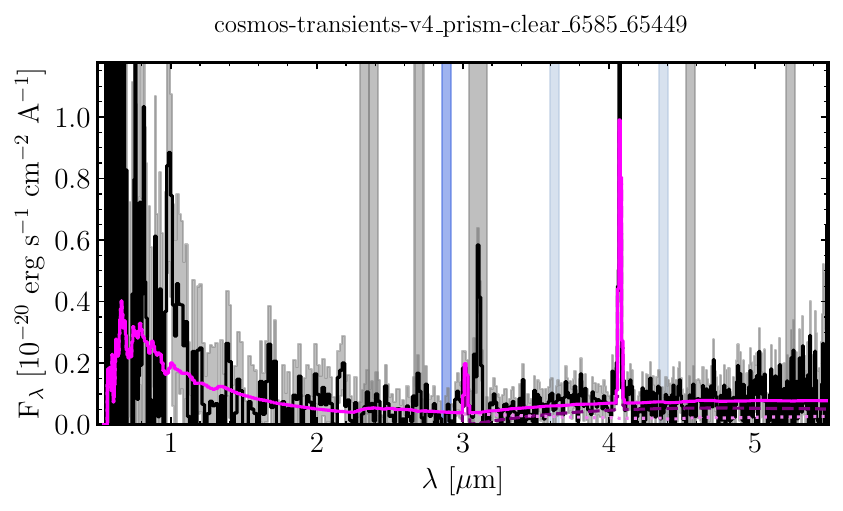}
\includegraphics[width=0.23\linewidth]{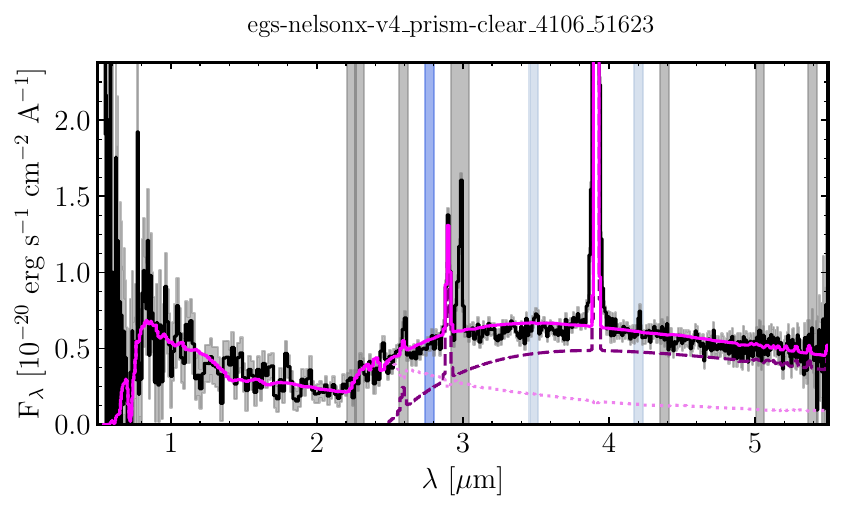}
\includegraphics[width=0.23\linewidth]{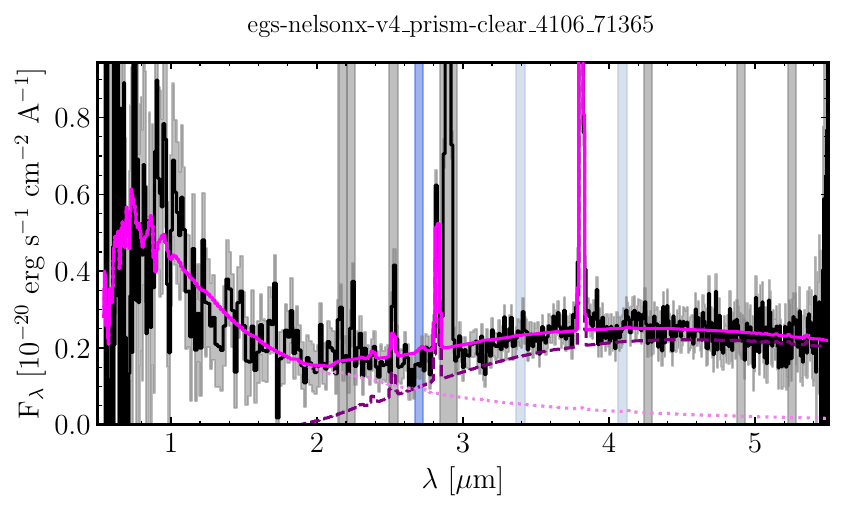}
\includegraphics[width=0.23\linewidth]{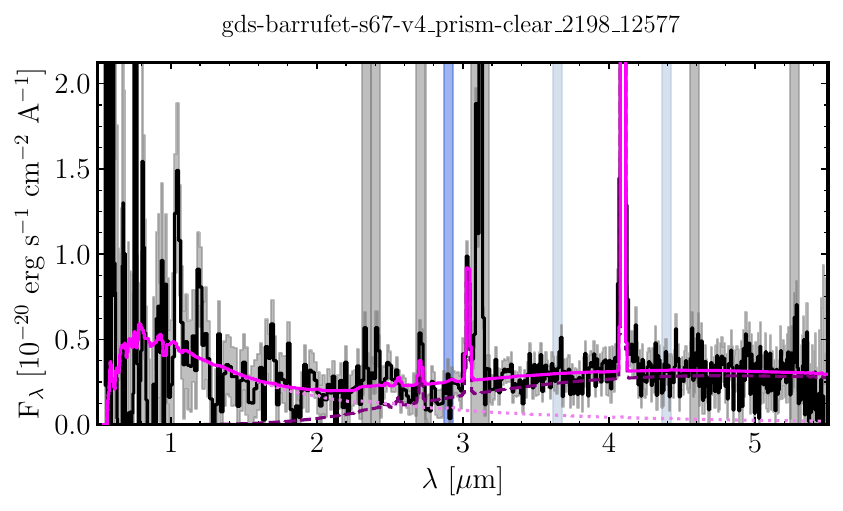}

\includegraphics[width=0.23\linewidth]{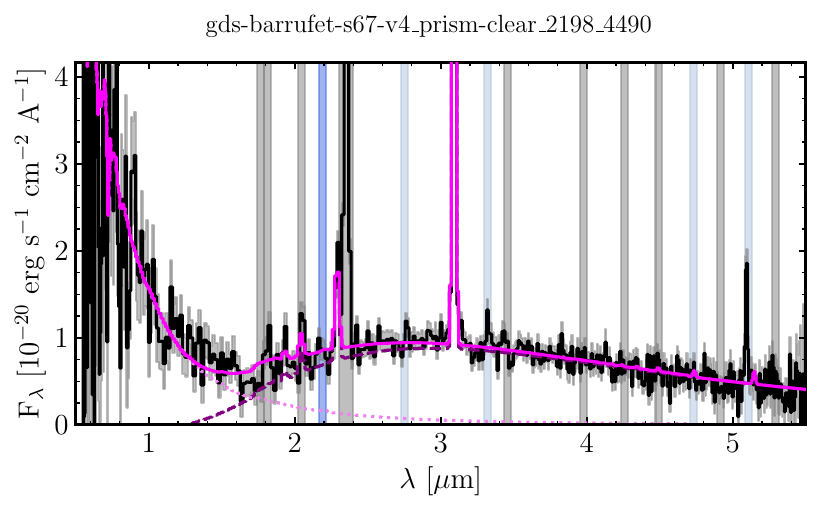}
\includegraphics[width=0.23\linewidth]{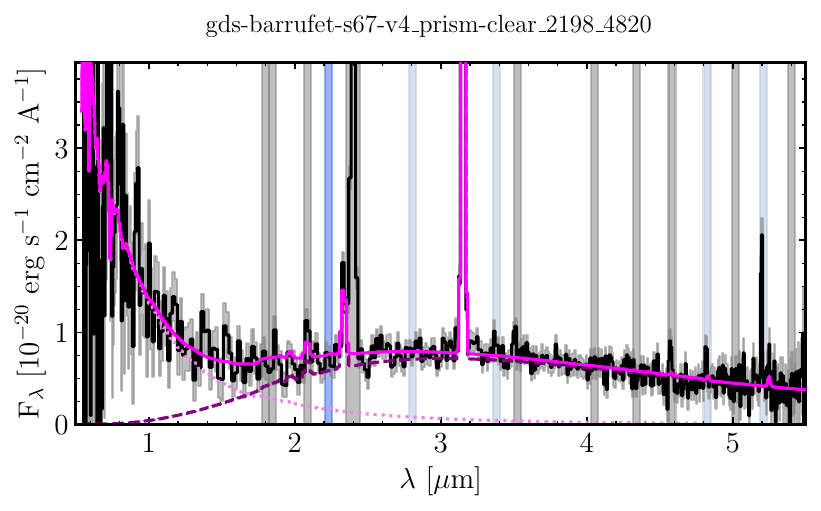}
\includegraphics[width=0.23\linewidth]{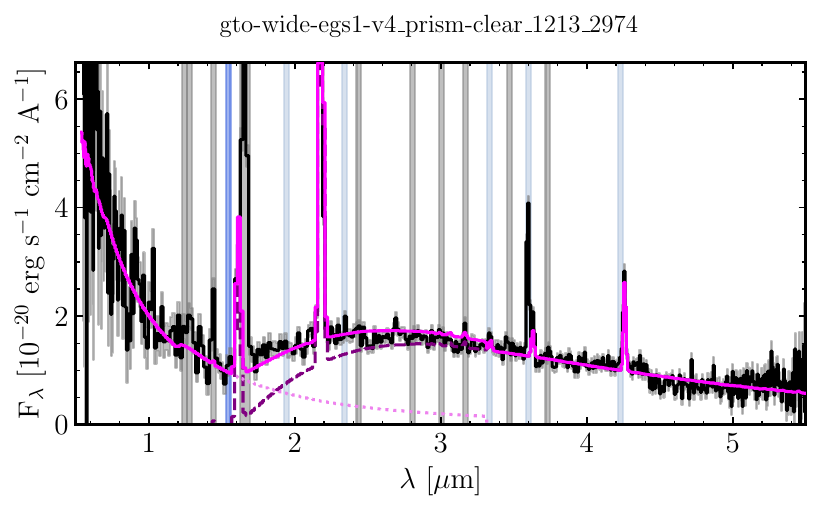}
\includegraphics[width=0.23\linewidth]{FIGURES/models/gto-wide-uds11-v4_prism-clear_1215_4994.spec.pdf}

\includegraphics[width=0.23\linewidth]{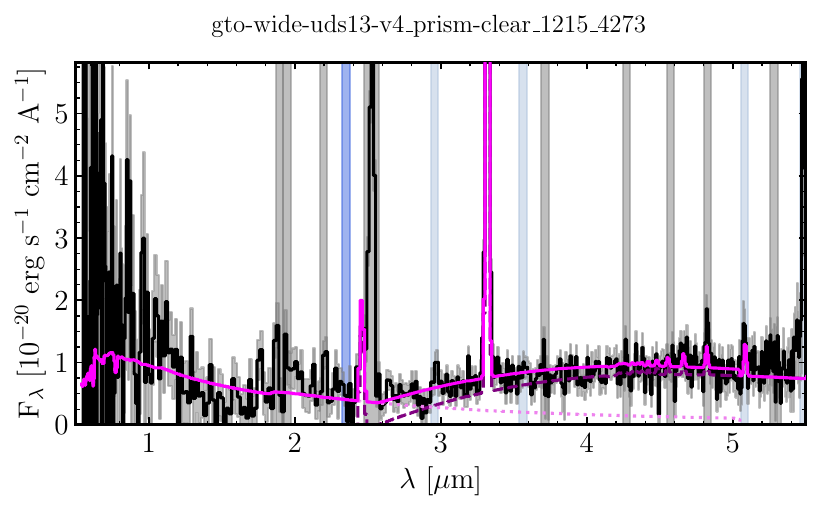}
\includegraphics[width=0.23\linewidth]{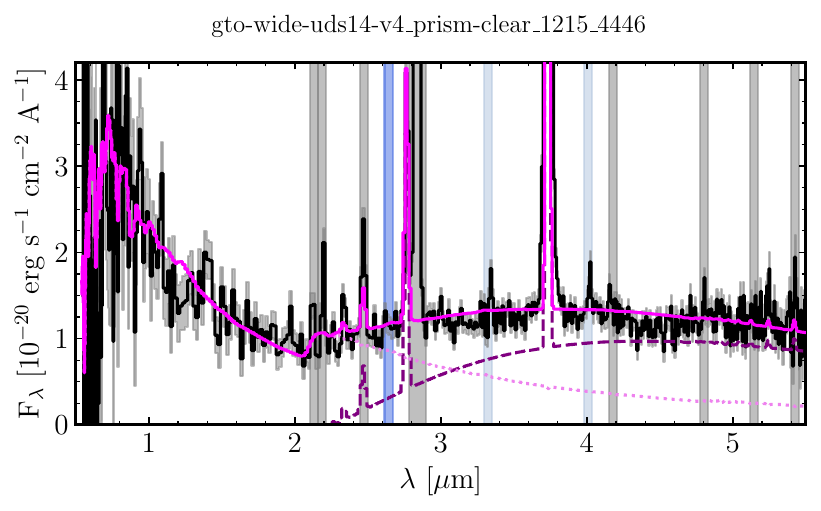}
\includegraphics[width=0.23\linewidth]{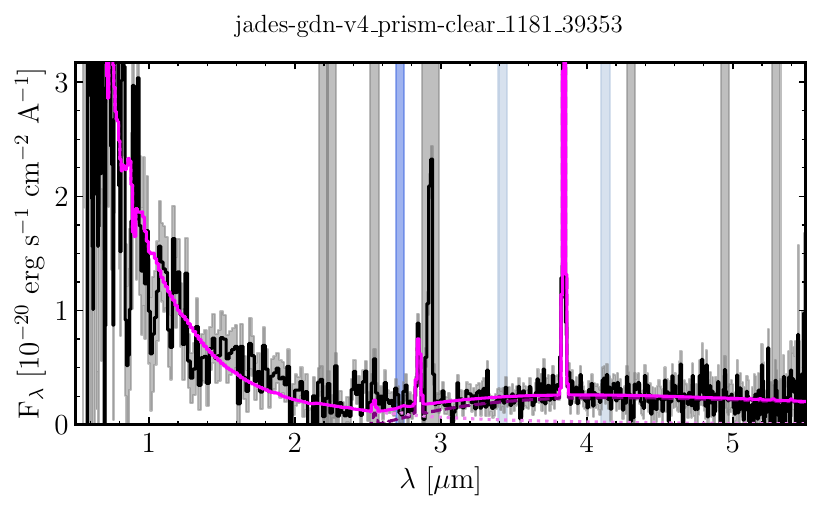}
\includegraphics[width=0.23\linewidth]{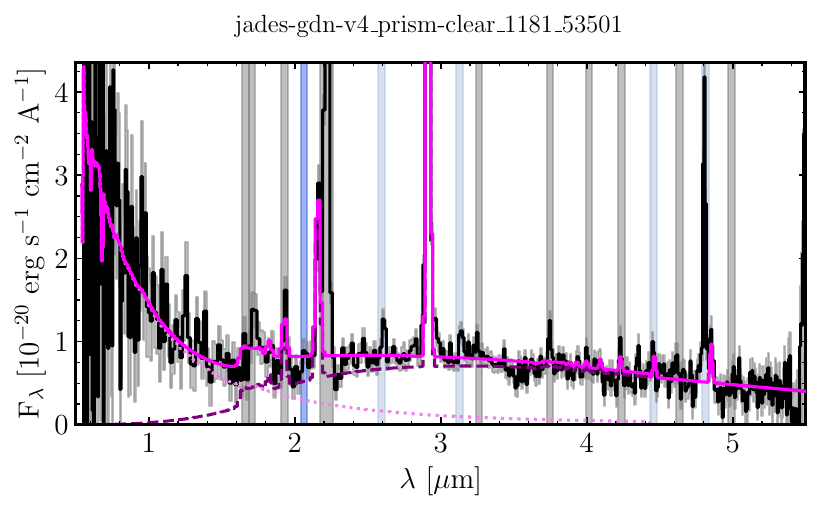}

\includegraphics[width=0.23\linewidth]{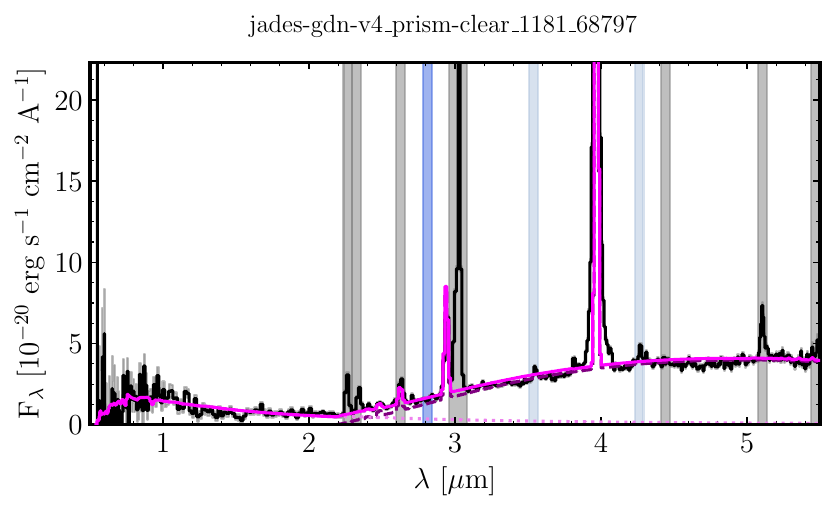}
\includegraphics[width=0.23\linewidth]{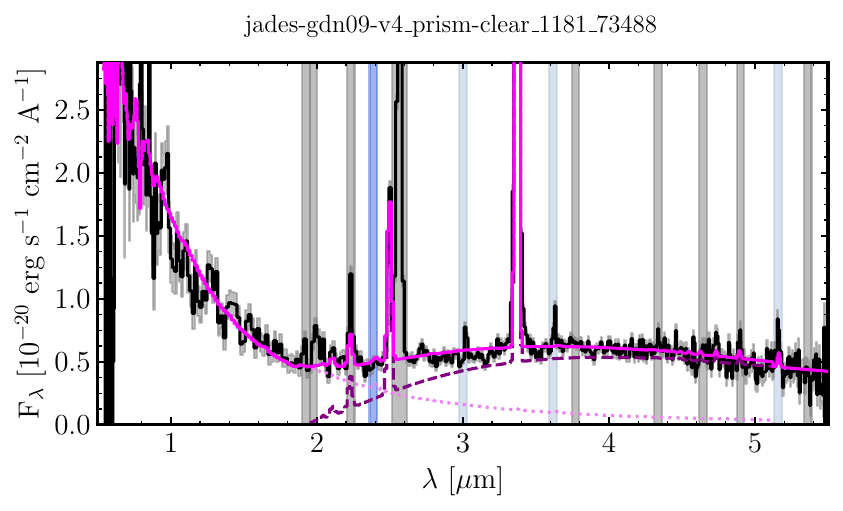}
\includegraphics[width=0.23\linewidth]{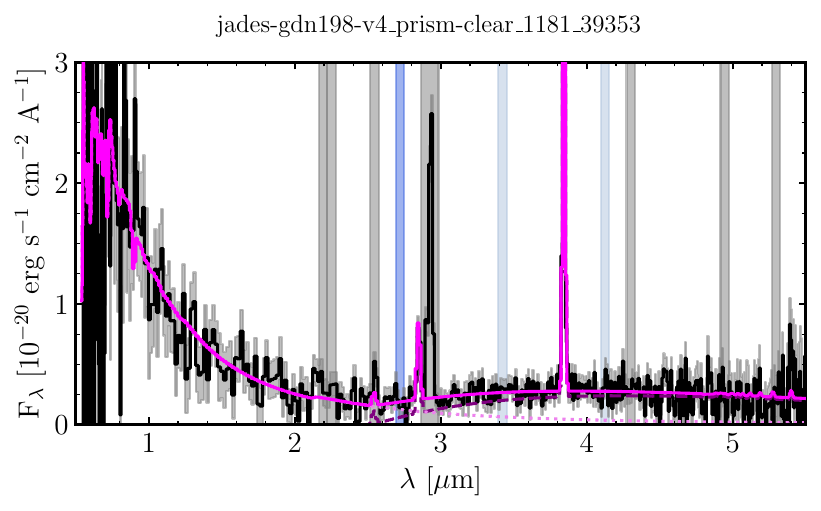}
\includegraphics[width=0.23\linewidth]{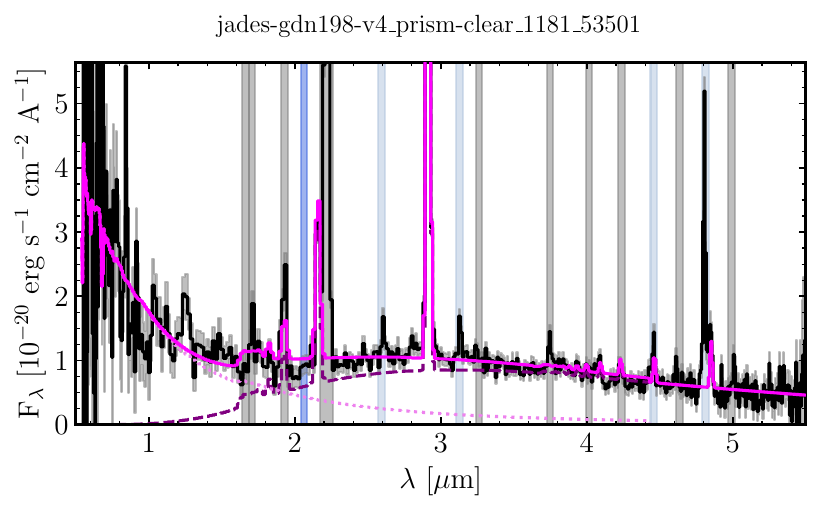}

\caption{Best-fitting models for the LRDs in the sample by \citet{DeGraaff_25b}. The colour scheme is the same as that employed in Fig. \ref{fig:examples}.}
\label{fig:spec_all}
\end{figure*}

\begin{figure*}[p]
\addtocounter{figure}{-1}\centering

\includegraphics[width=0.23\linewidth]{FIGURES/models/jades-gdn198-v4_prism-clear_1181_68797.spec.pdf}
\includegraphics[width=0.23\linewidth]{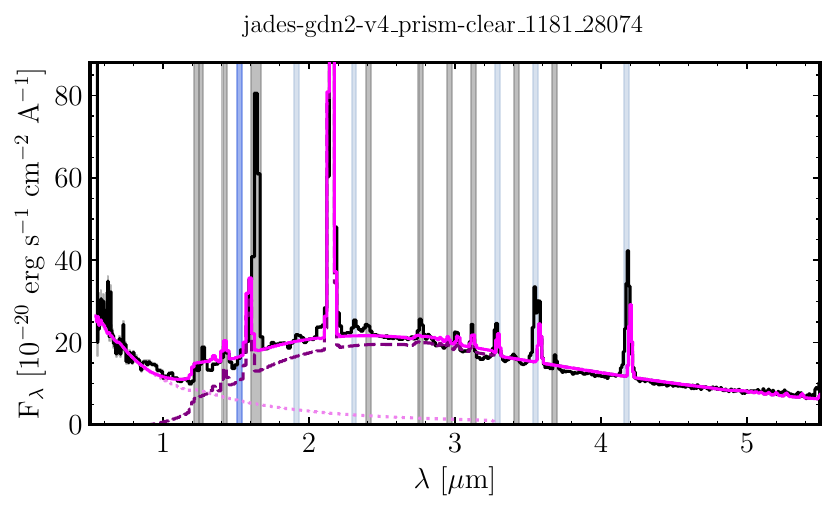}
\includegraphics[width=0.23\linewidth]{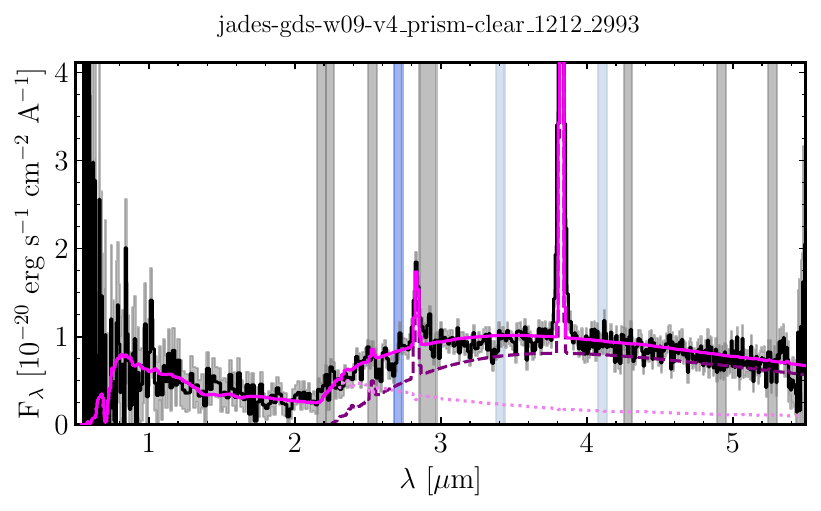}
\includegraphics[width=0.23\linewidth]{FIGURES/models/jades-gds-wide-v4_prism-clear_1180_12402.spec.pdf}

\includegraphics[width=0.23\linewidth]{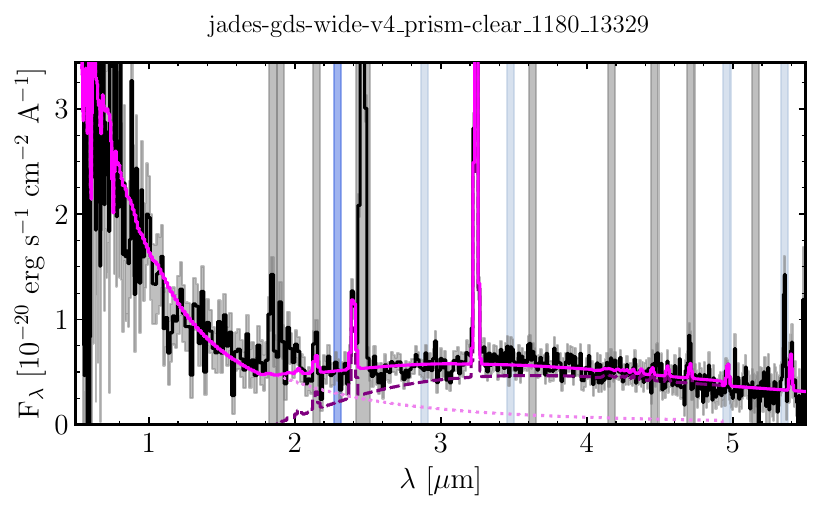}
\includegraphics[width=0.23\linewidth]{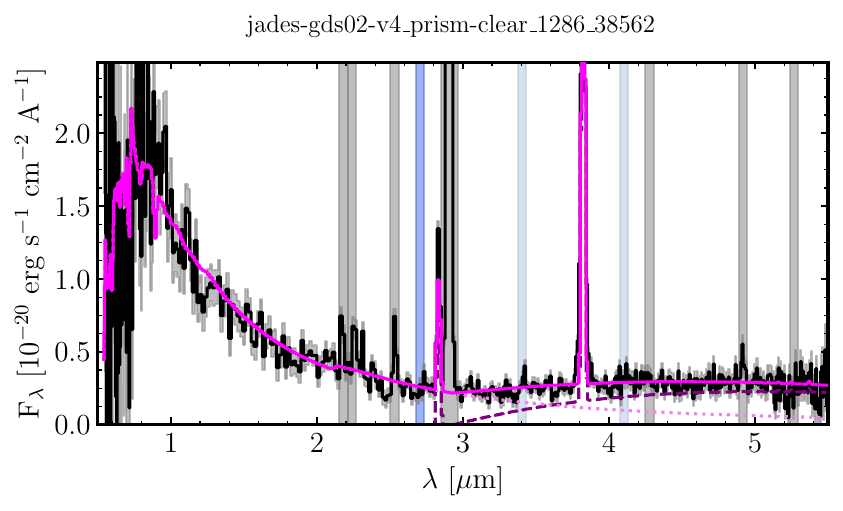}
\includegraphics[width=0.23\linewidth]{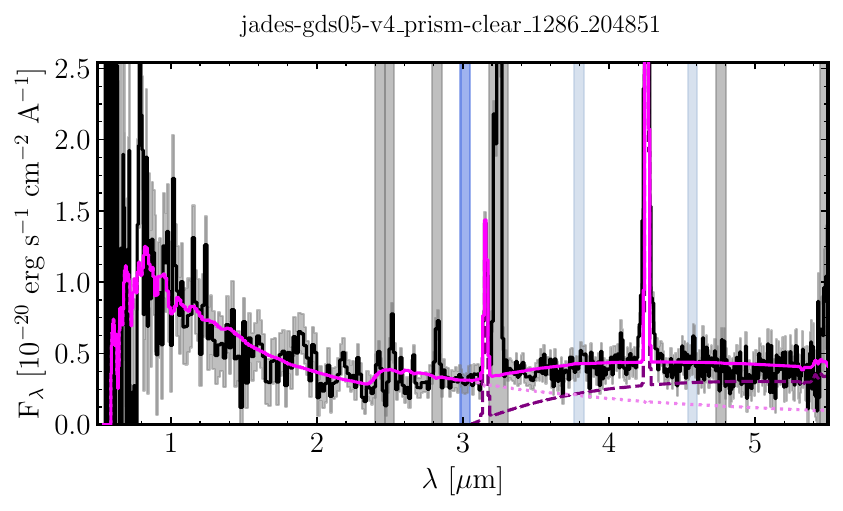}
\includegraphics[width=0.23\linewidth]{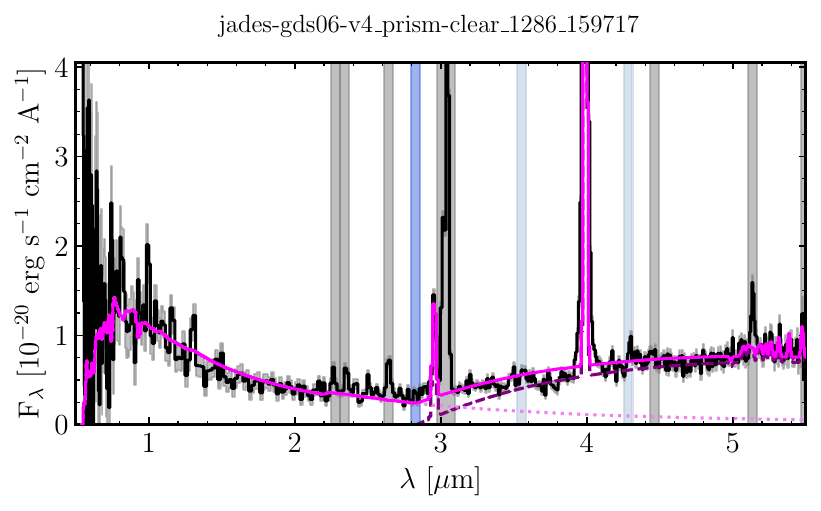}

\includegraphics[width=0.23\linewidth]{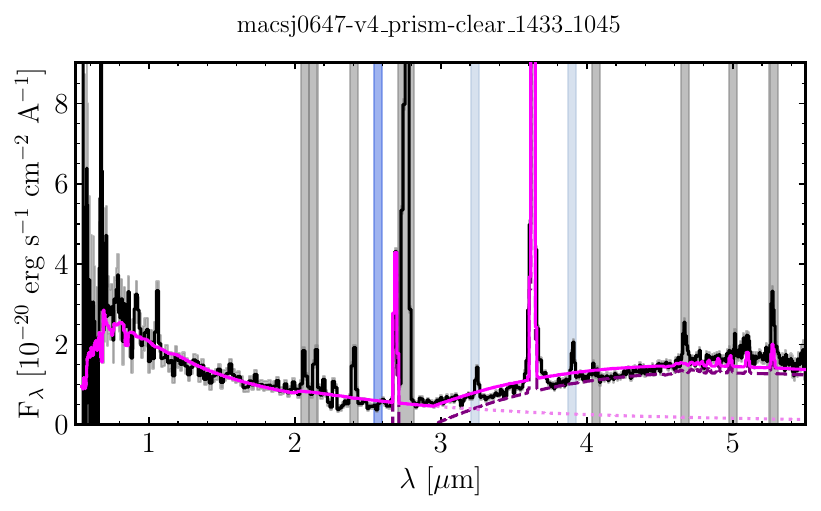}
\includegraphics[width=0.23\linewidth]{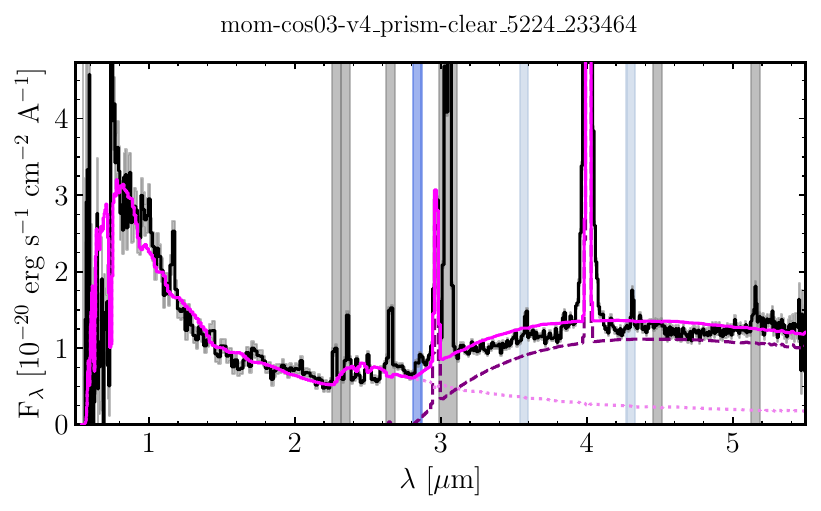}
\includegraphics[width=0.23\linewidth]{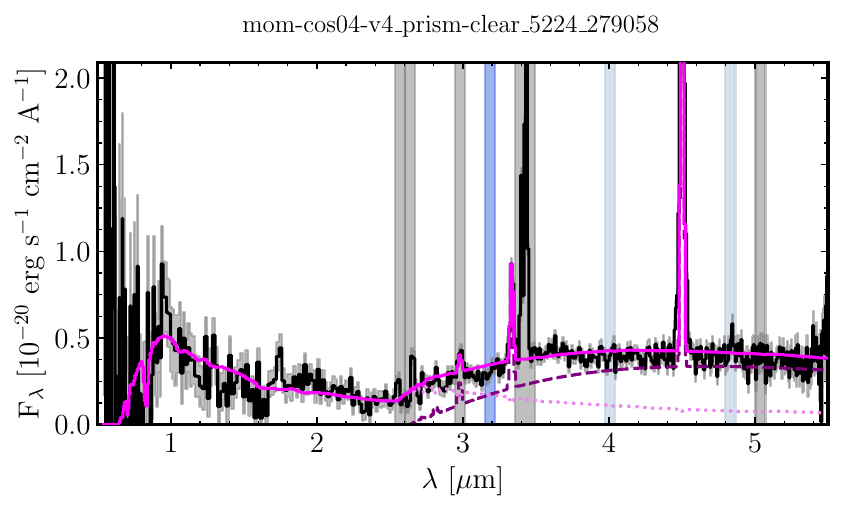}
\includegraphics[width=0.23\linewidth]{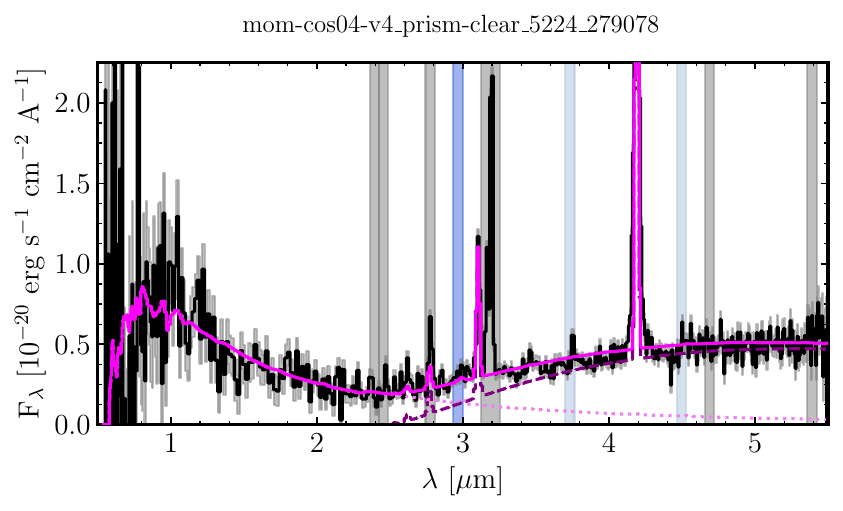}

\includegraphics[width=0.23\linewidth]{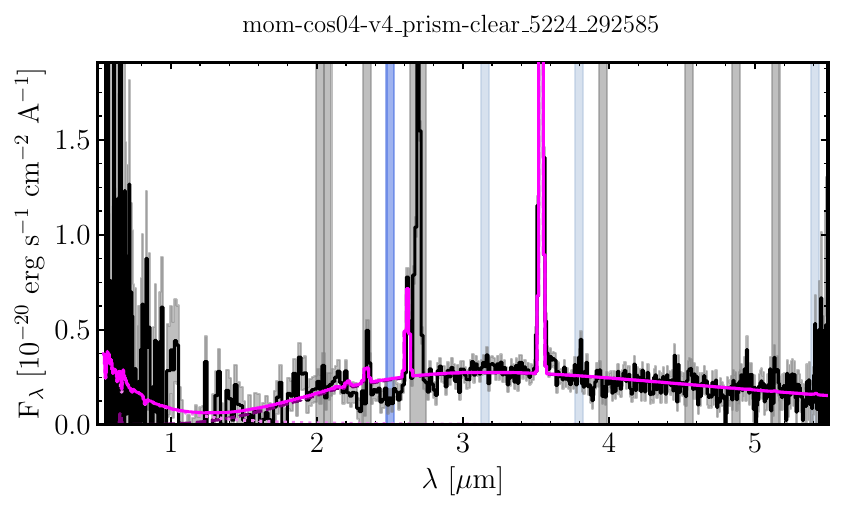}
\includegraphics[width=0.23\linewidth]{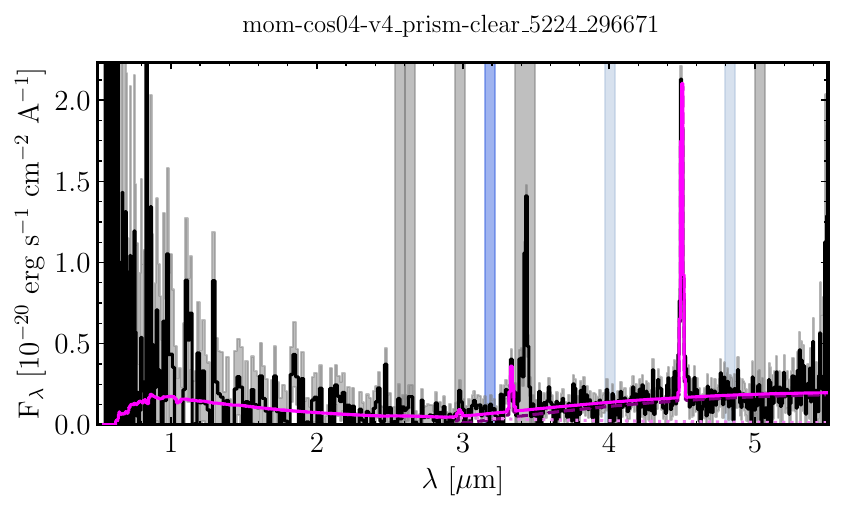}
\includegraphics[width=0.23\linewidth]{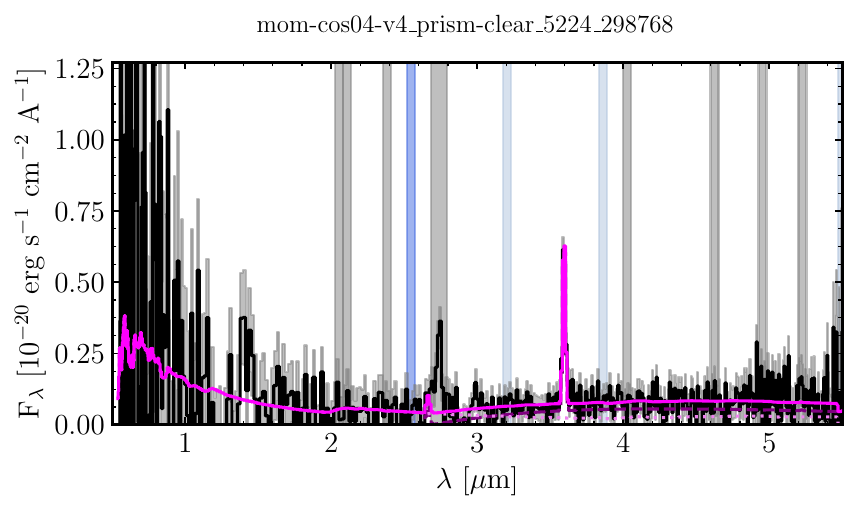}
\includegraphics[width=0.23\linewidth]{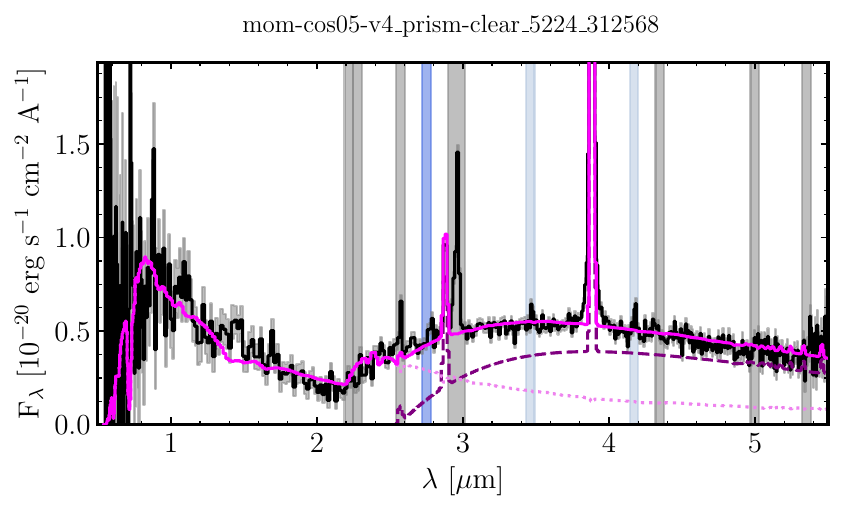}

\includegraphics[width=0.23\linewidth]{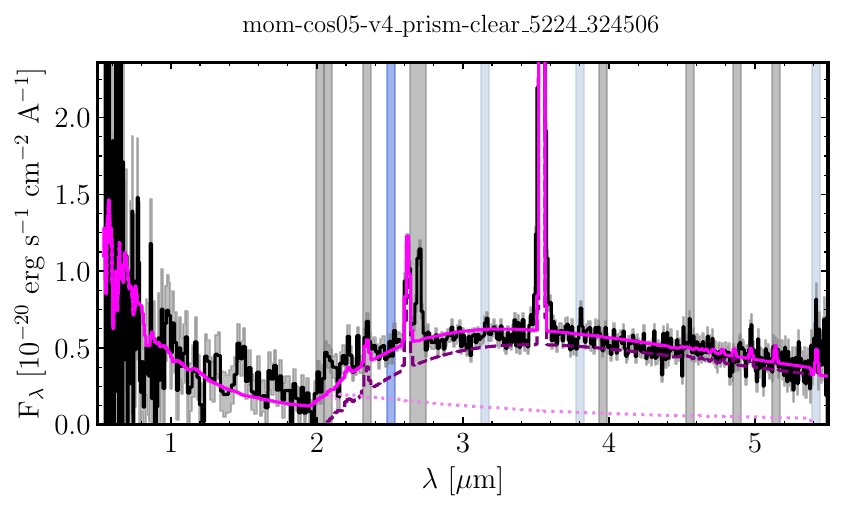}
\includegraphics[width=0.23\linewidth]{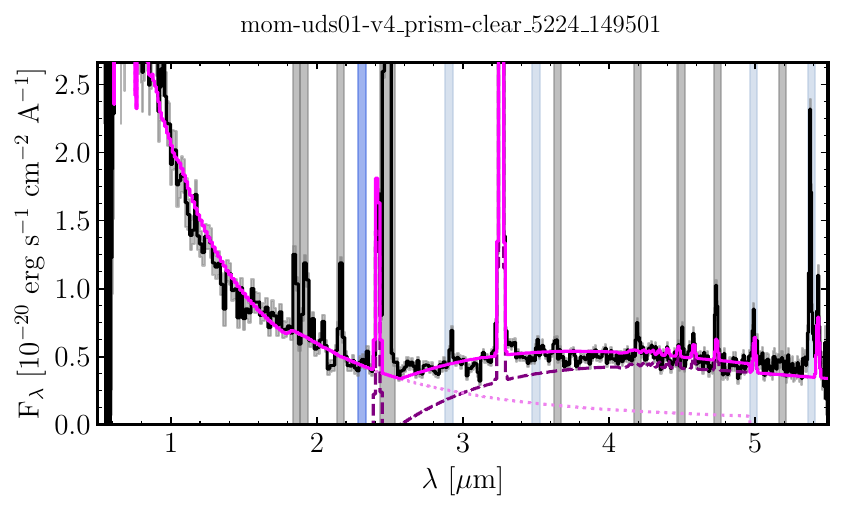}
\includegraphics[width=0.23\linewidth]{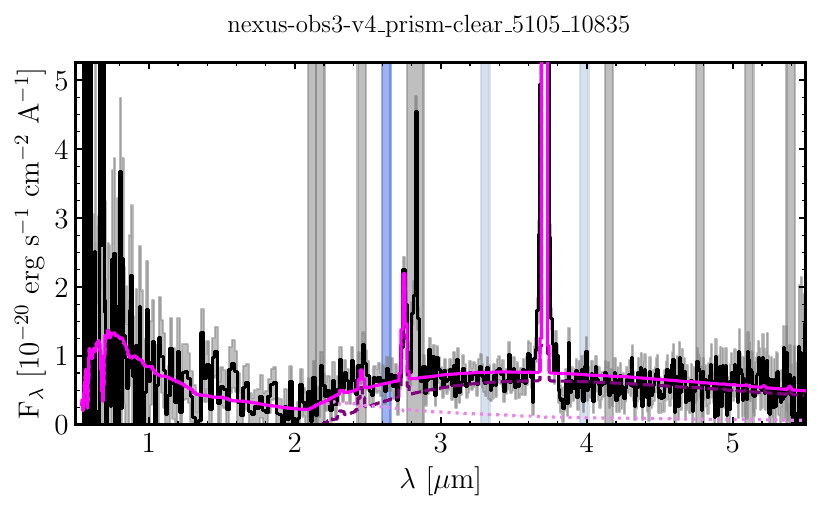}
\includegraphics[width=0.23\linewidth]{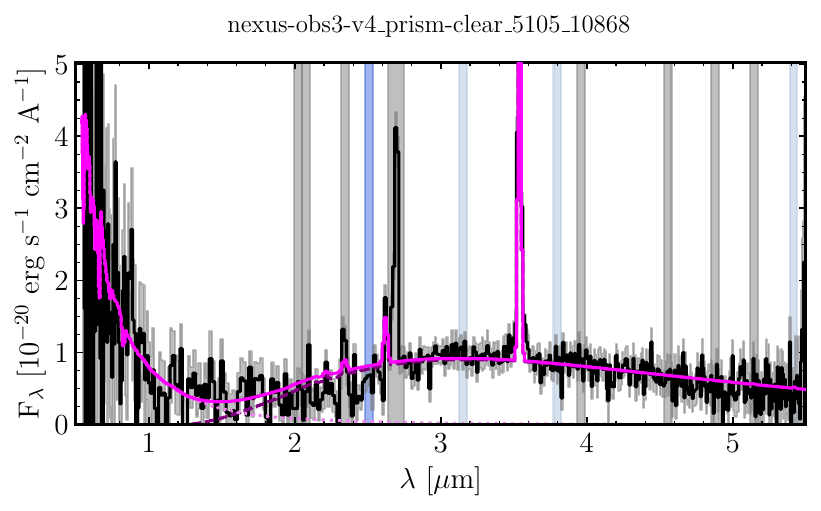}

\includegraphics[width=0.23\linewidth]{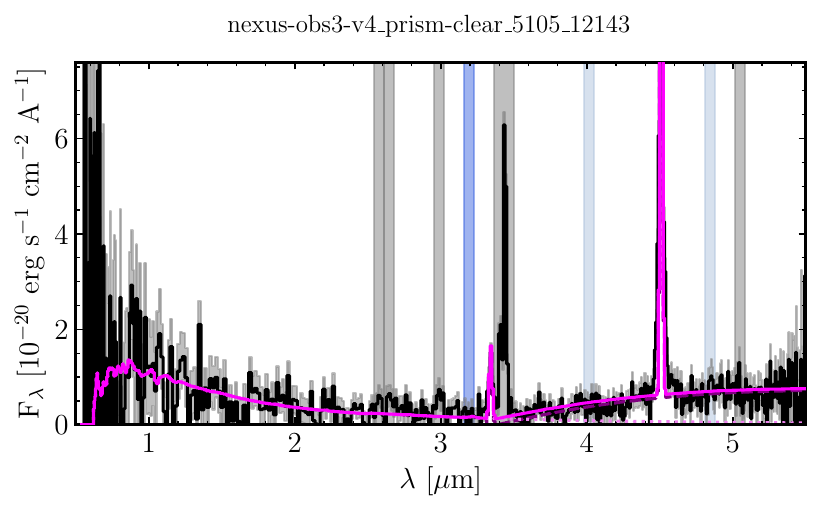}
\includegraphics[width=0.23\linewidth]{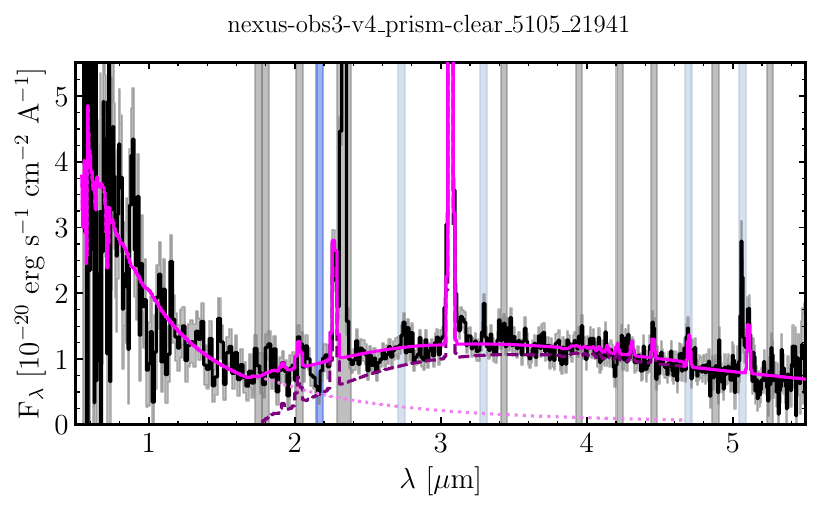}
\includegraphics[width=0.23\linewidth]{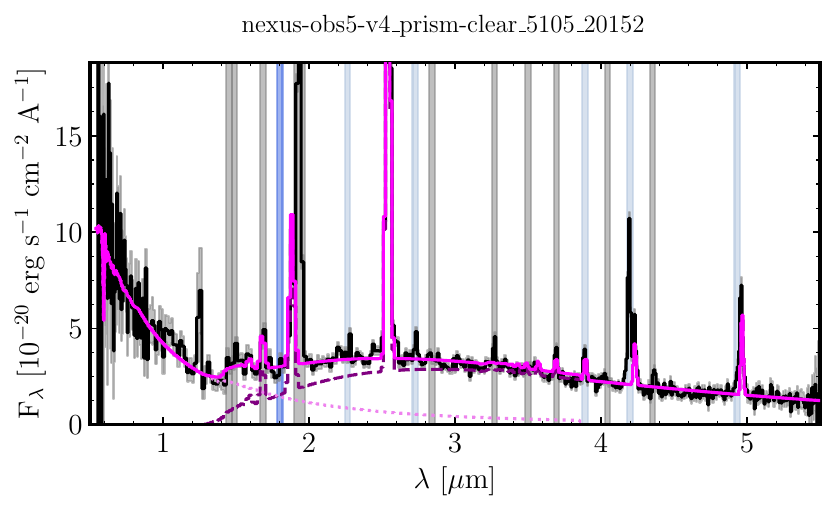}
\includegraphics[width=0.23\linewidth]{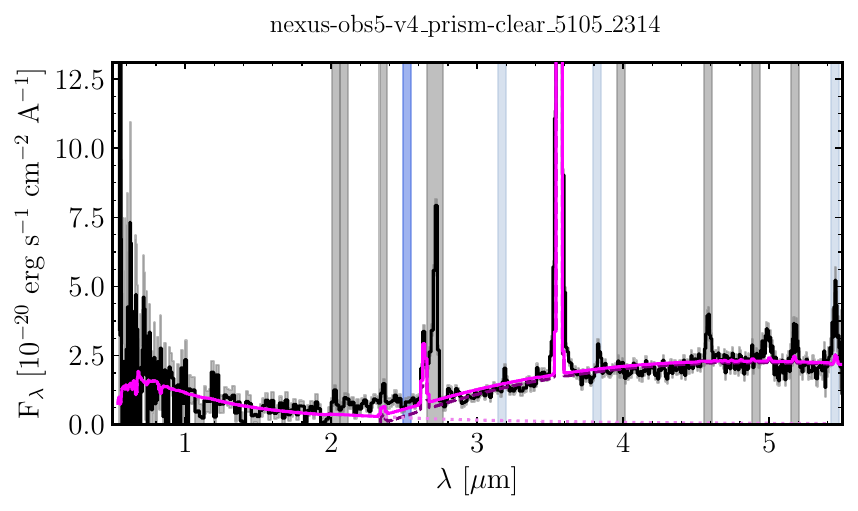}

\includegraphics[width=0.23\linewidth]{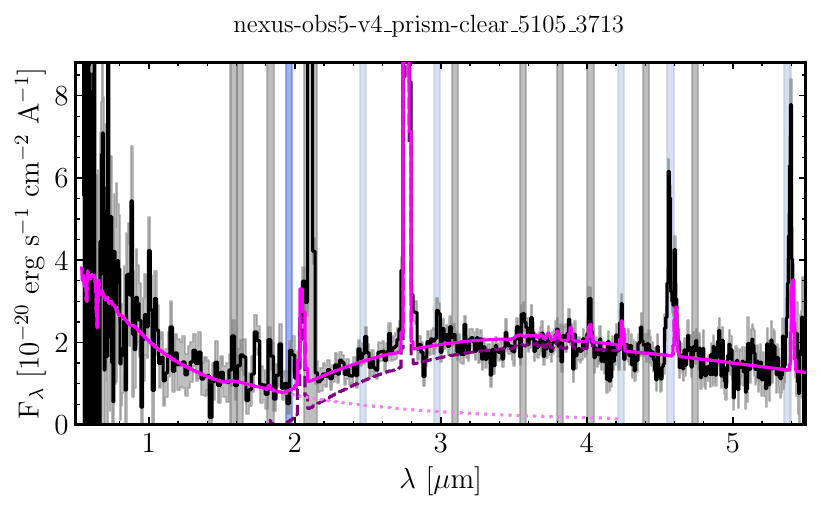}
\includegraphics[width=0.23\linewidth]{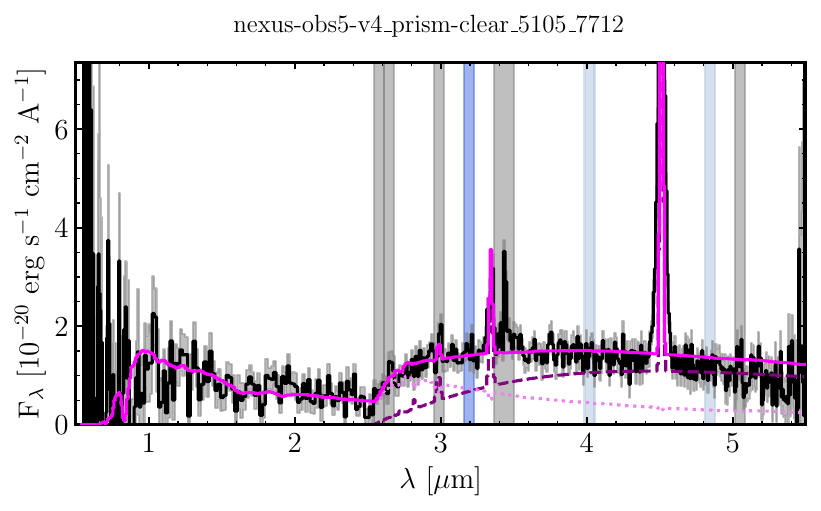}
\includegraphics[width=0.23\linewidth]{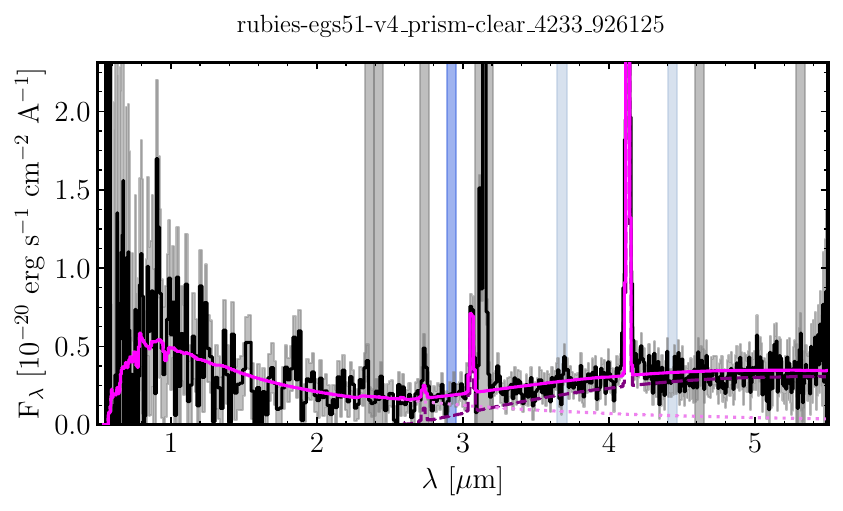}
\includegraphics[width=0.23\linewidth]{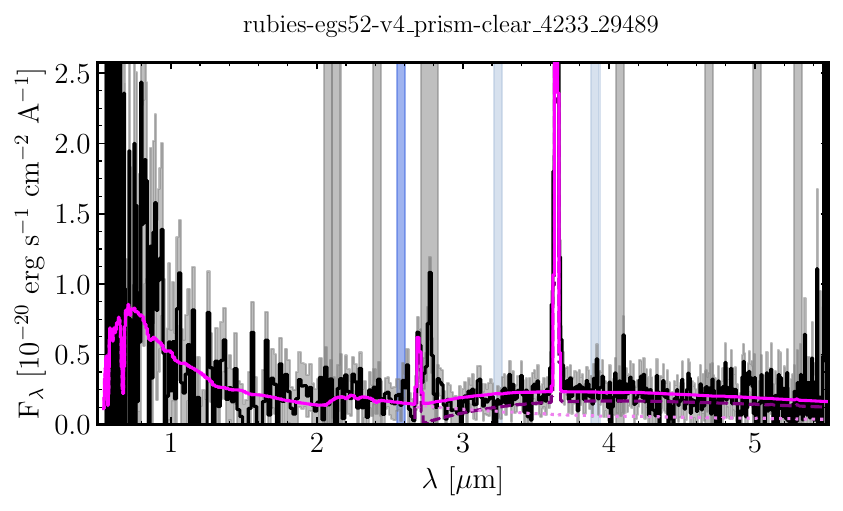}

\includegraphics[width=0.23\linewidth]{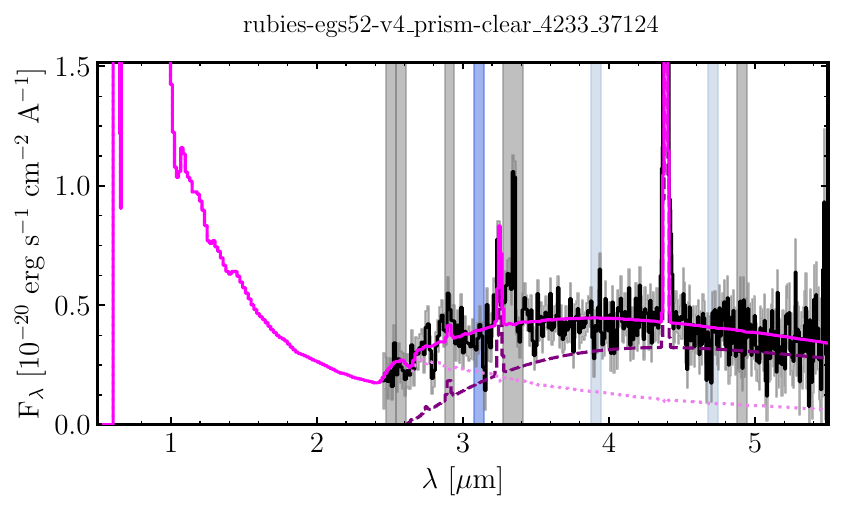}
\includegraphics[width=0.23\linewidth]{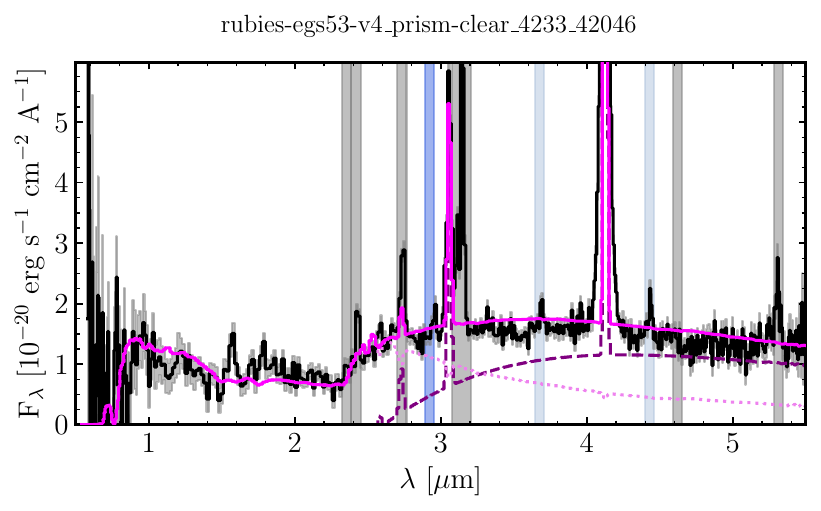}
\includegraphics[width=0.23\linewidth]{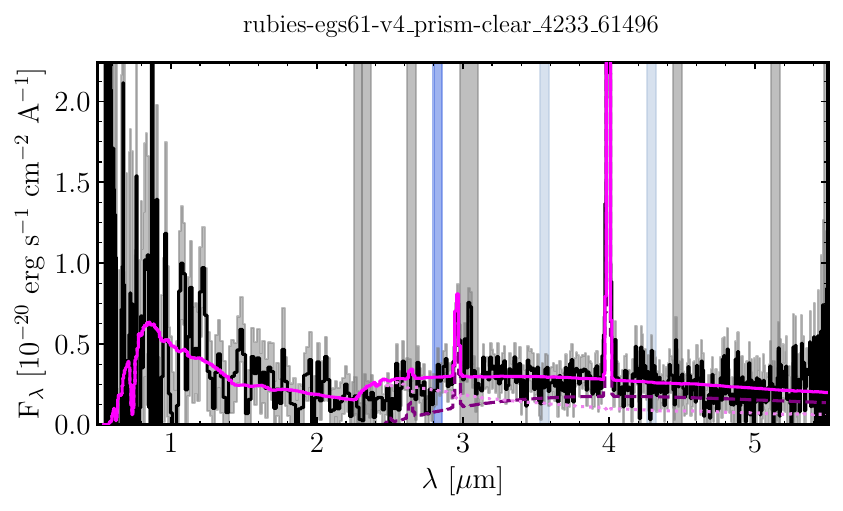}
\includegraphics[width=0.23\linewidth]{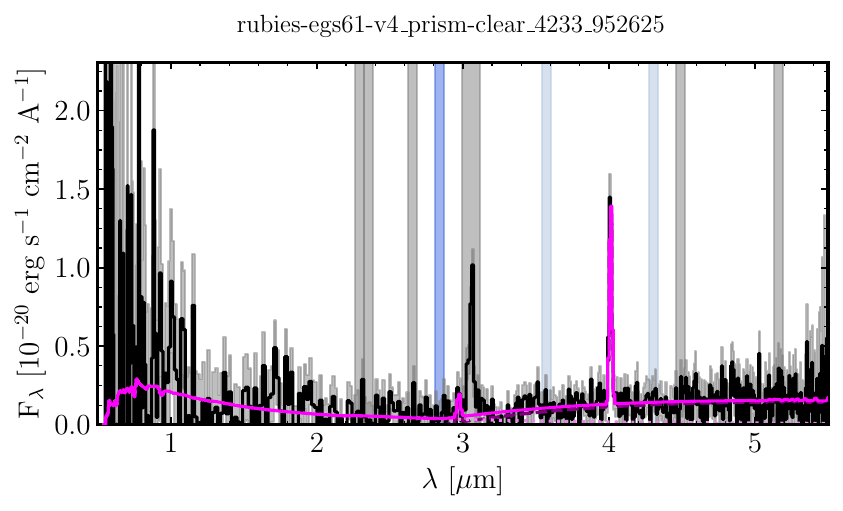}

\caption[]{Continued.}
\end{figure*}

\begin{figure*}[p]
\addtocounter{figure}{-1}\centering

\includegraphics[width=0.23\linewidth]{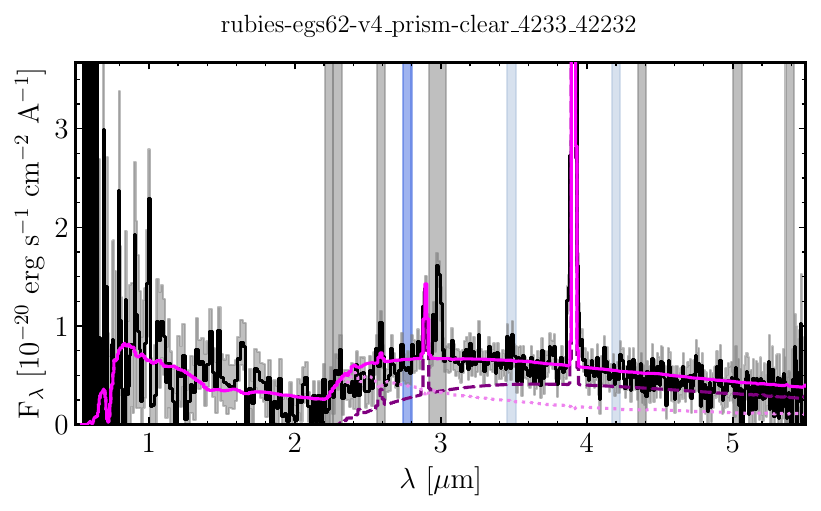}
\includegraphics[width=0.23\linewidth]{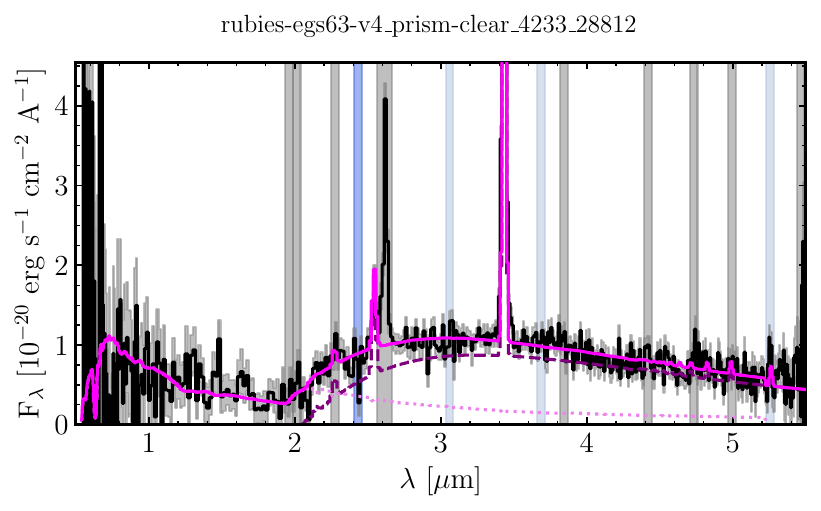}
\includegraphics[width=0.23\linewidth]{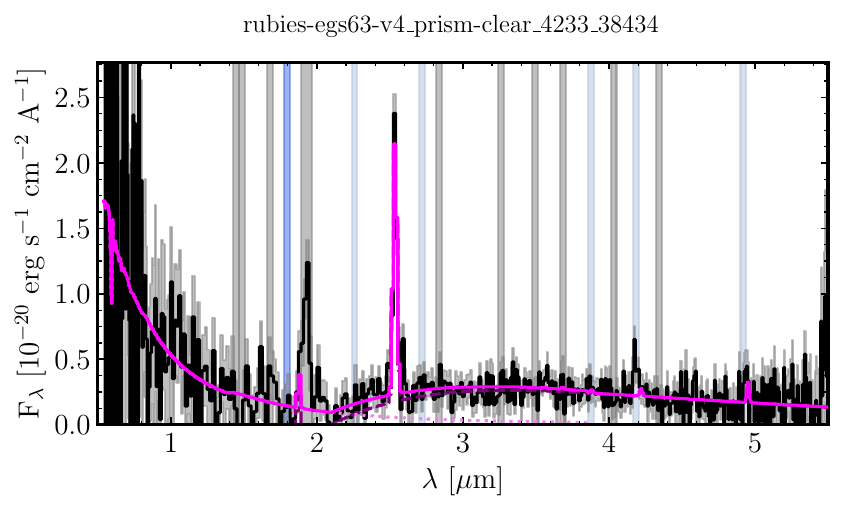}
\includegraphics[width=0.23\linewidth]{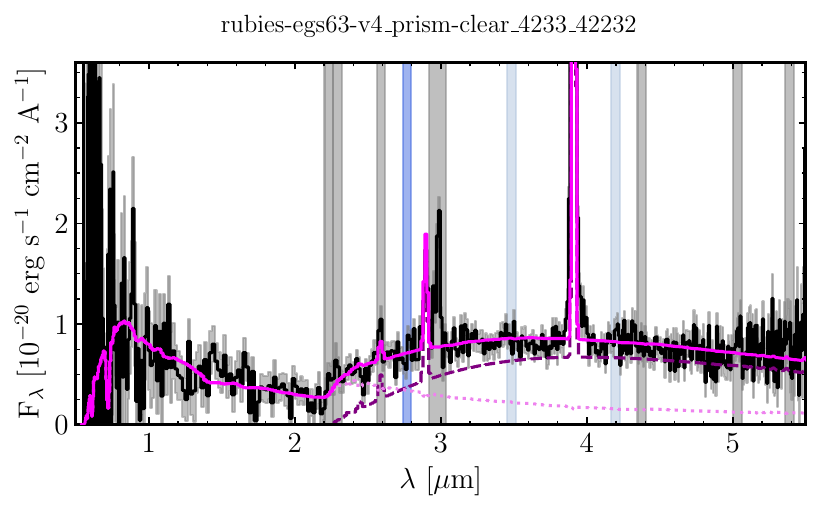}

\includegraphics[width=0.23\linewidth]{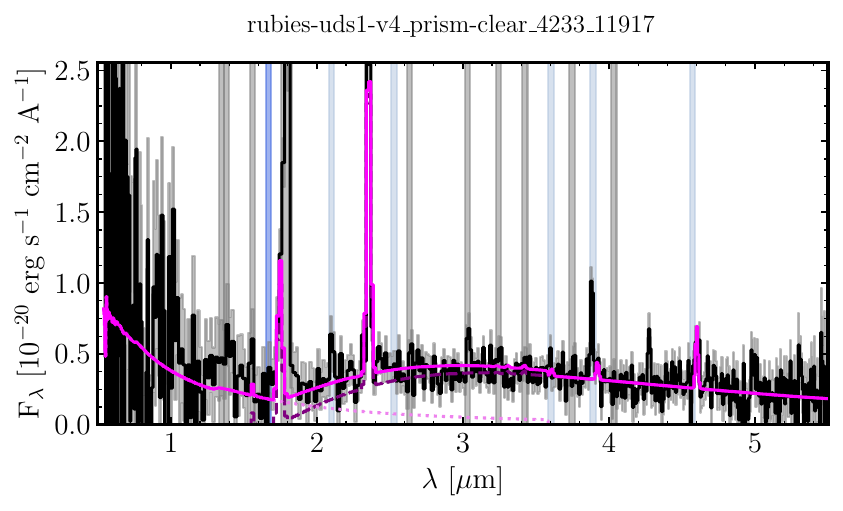}
\includegraphics[width=0.23\linewidth]{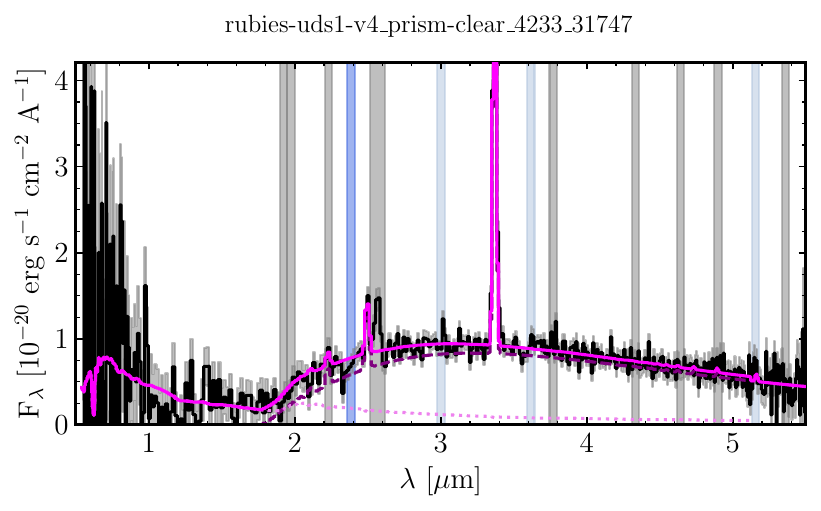}
\includegraphics[width=0.23\linewidth]{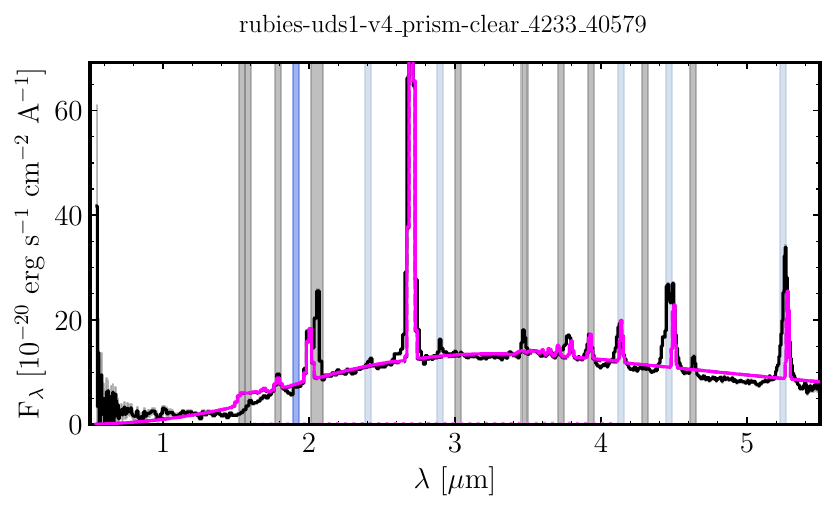}
\includegraphics[width=0.23\linewidth]{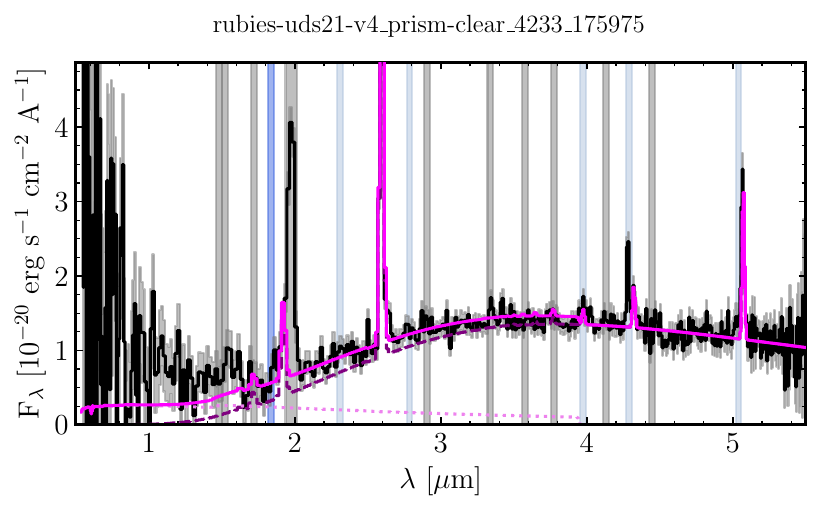}

\includegraphics[width=0.23\linewidth]{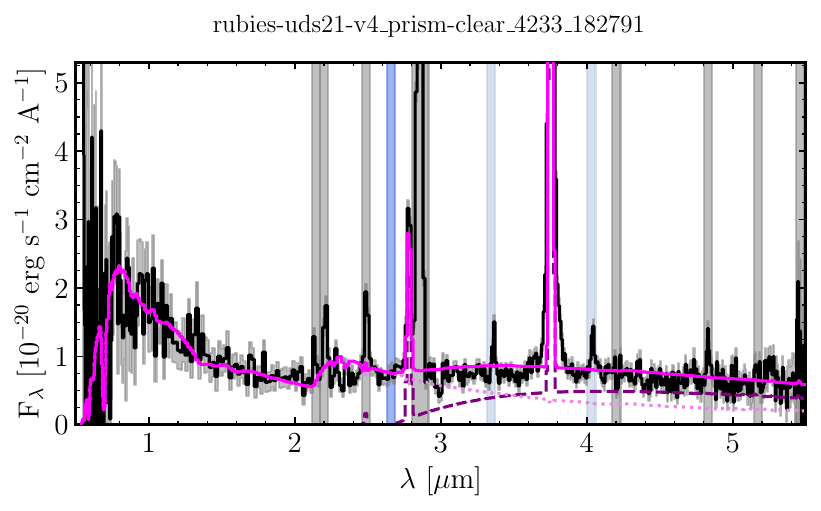}
\includegraphics[width=0.23\linewidth]{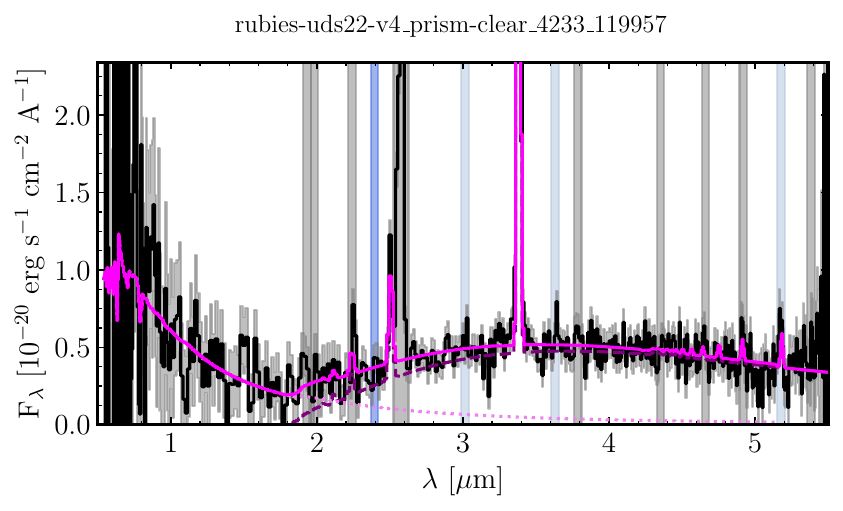}
\includegraphics[width=0.23\linewidth]{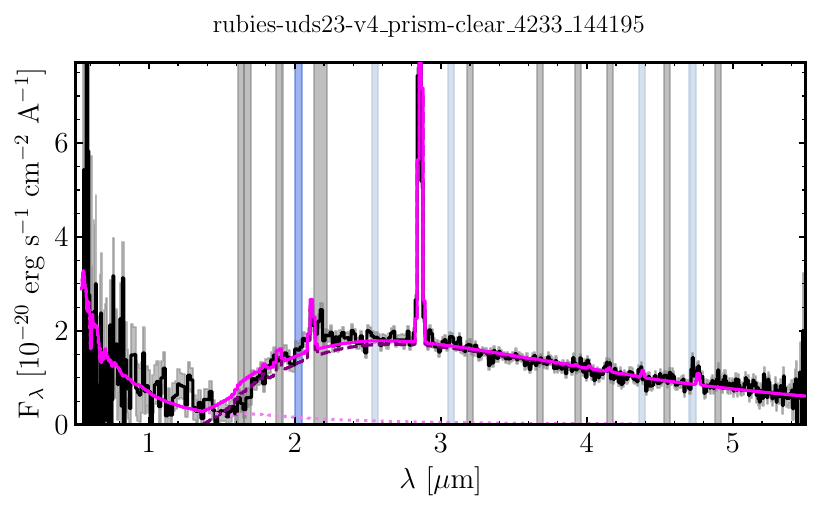}
\includegraphics[width=0.23\linewidth]{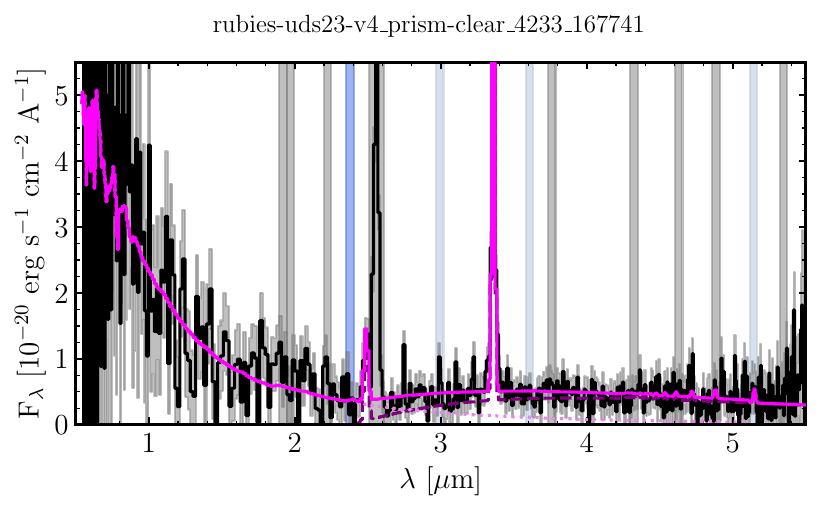}

\includegraphics[width=0.23\linewidth]{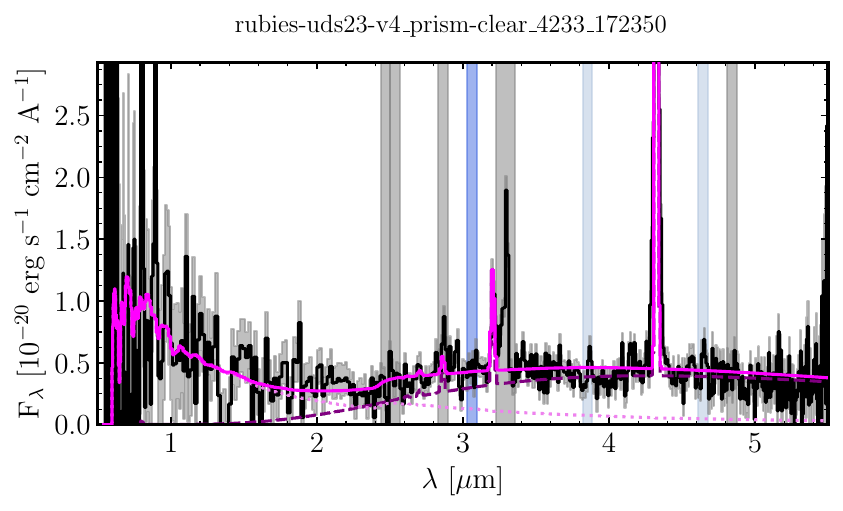}
\includegraphics[width=0.23\linewidth]{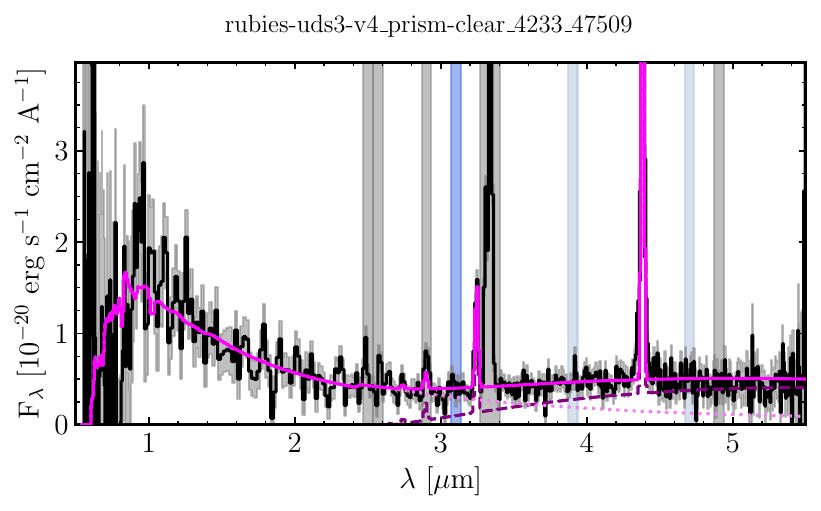}
\includegraphics[width=0.23\linewidth]{FIGURES/models/rubies-uds31-v4_prism-clear_4233_154183.spec.pdf}
\includegraphics[width=0.23\linewidth]{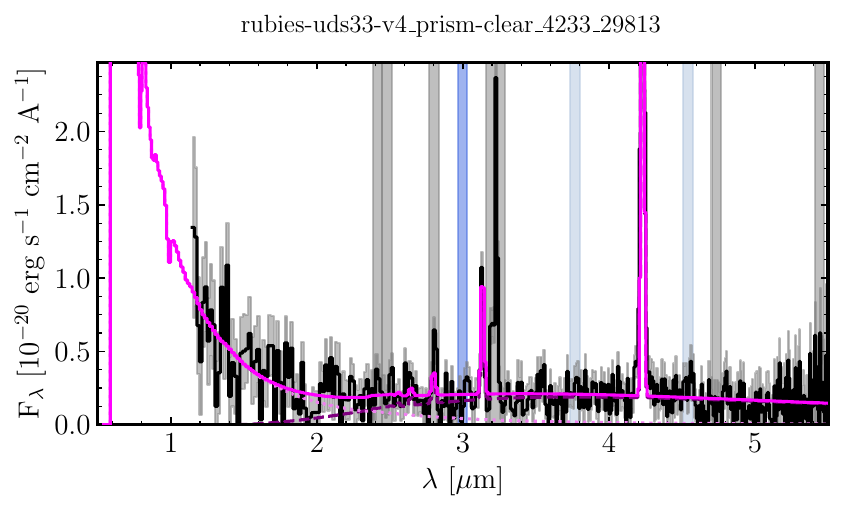}

\includegraphics[width=0.23\linewidth]{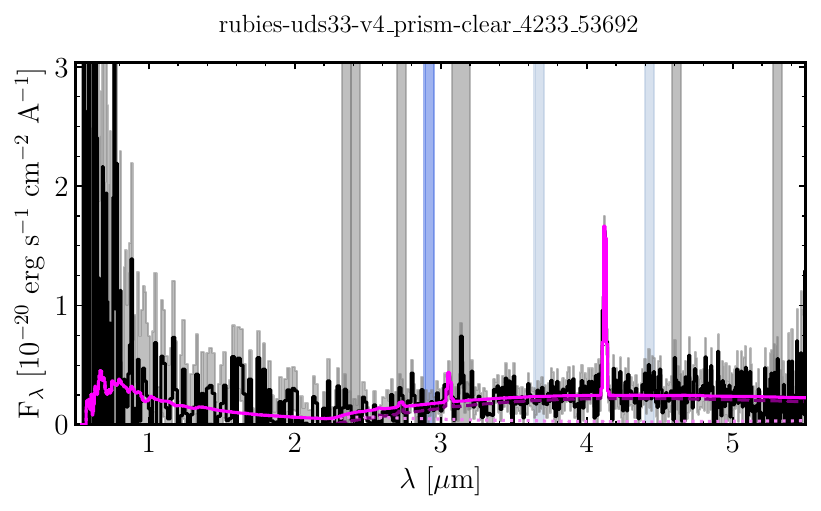}
\includegraphics[width=0.23\linewidth]{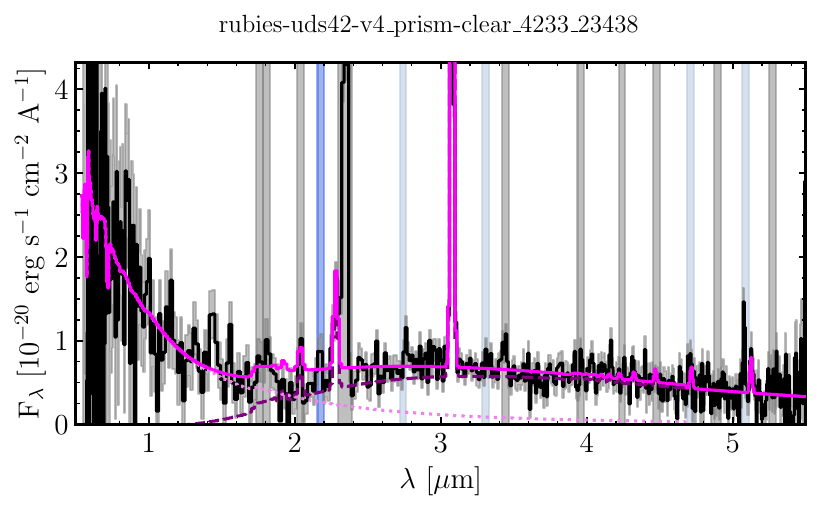}
\includegraphics[width=0.23\linewidth]{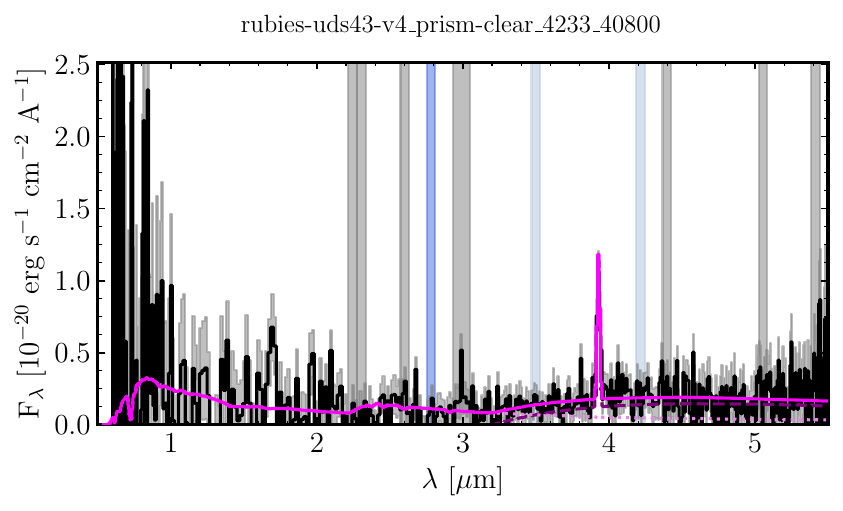}
\includegraphics[width=0.23\linewidth]{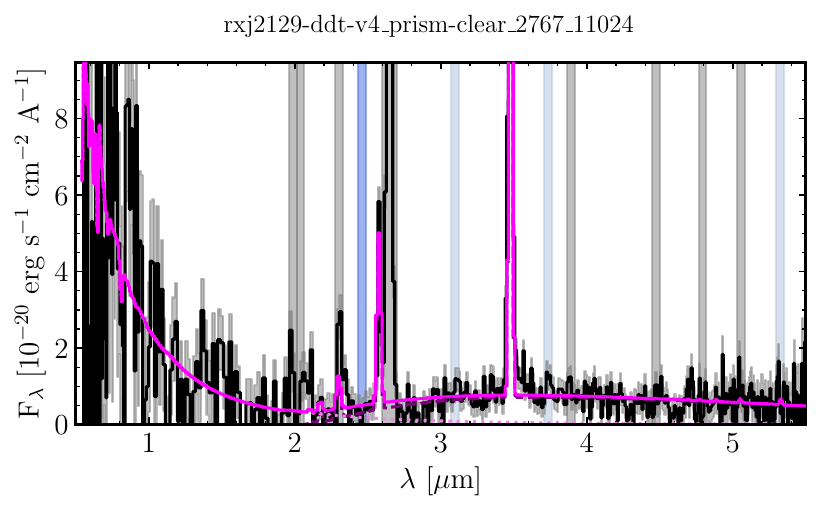}

\includegraphics[width=0.23\linewidth]{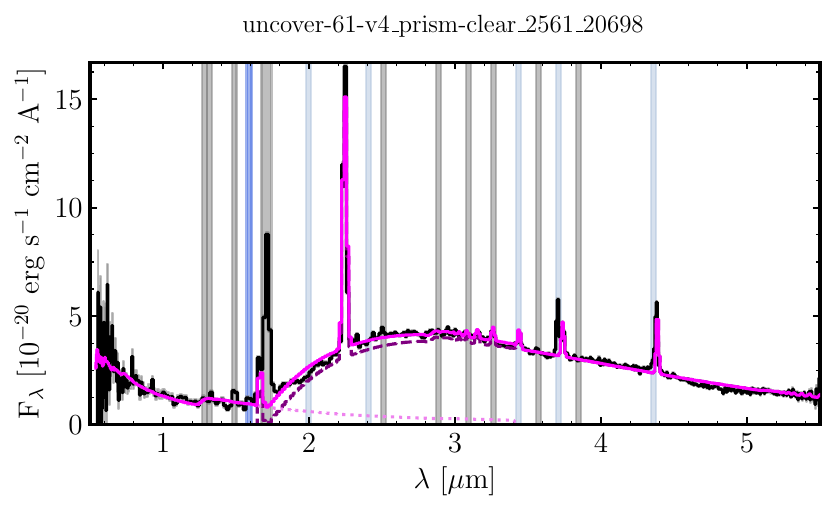}
\includegraphics[width=0.23\linewidth]{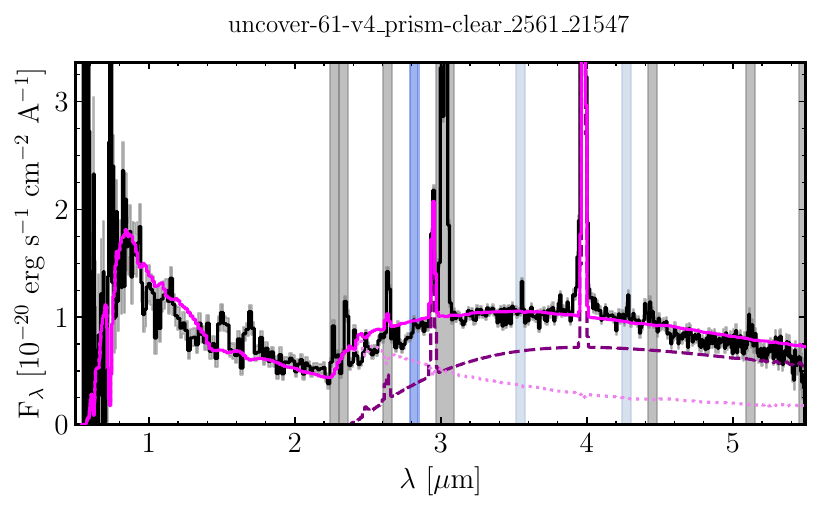}
\includegraphics[width=0.23\linewidth]{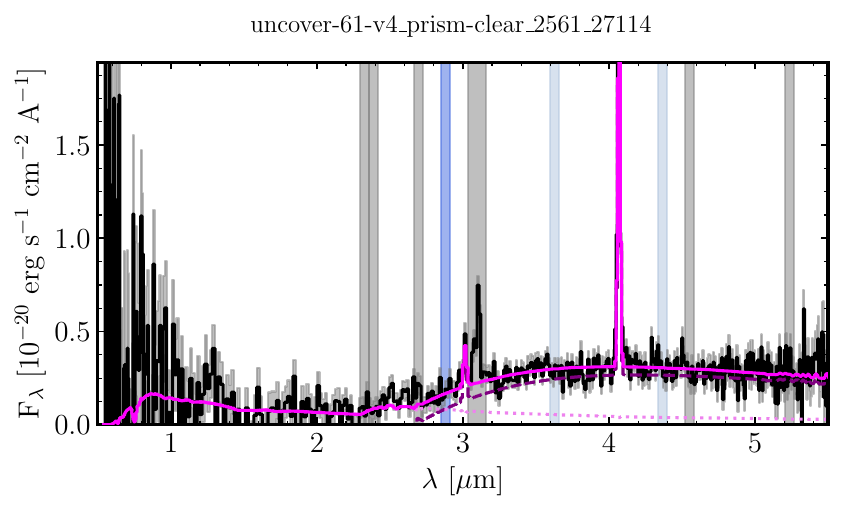}
\includegraphics[width=0.23\linewidth]{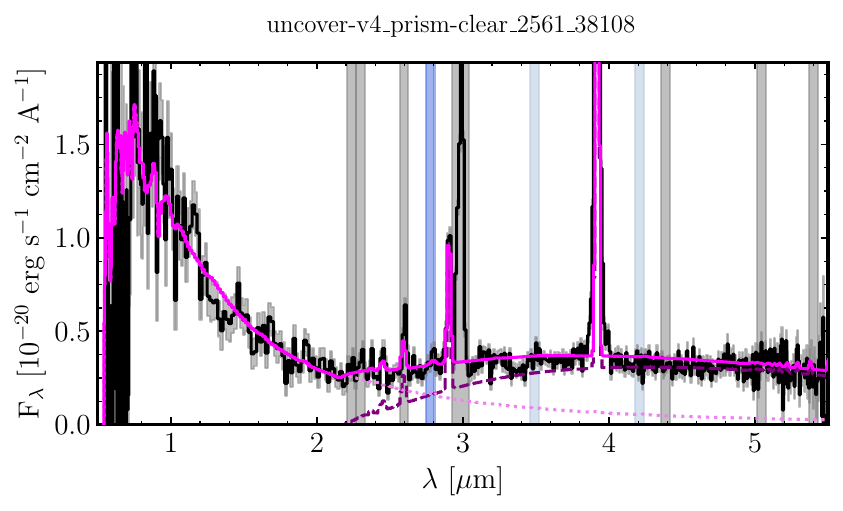}

\includegraphics[width=0.23\linewidth]{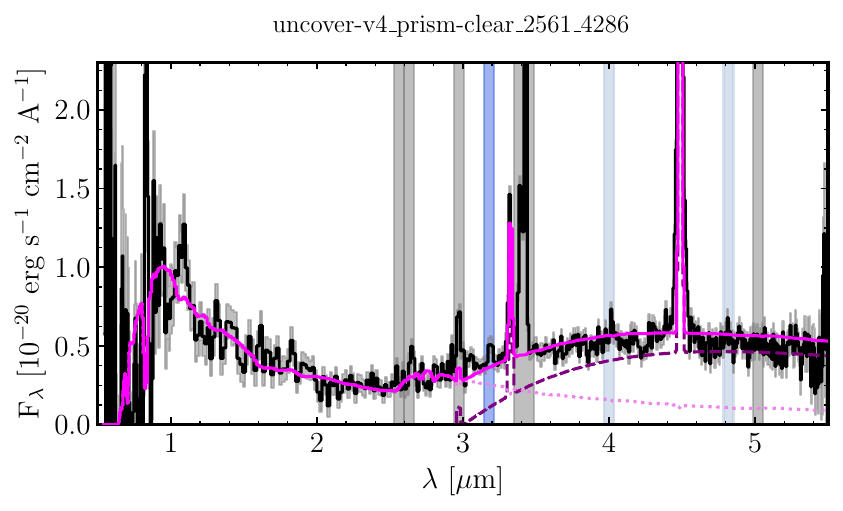}
\includegraphics[width=0.23\linewidth]{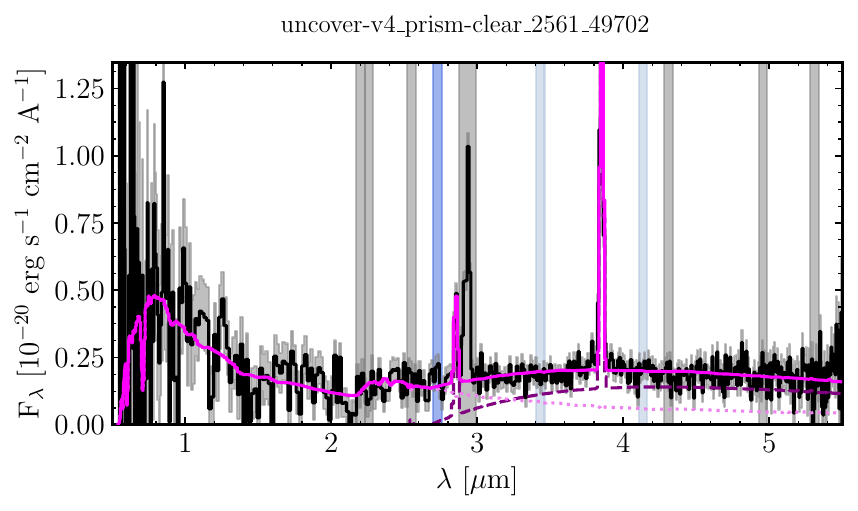}
\caption[]{Continued.}
\end{figure*}

\end{appendix}

\end{document}